\documentclass[prc,showpacs,showkeys]{revtex4}
\usepackage{epsfig,graphicx,float,color}
\usepackage{amssymb}
\usepackage[normalem]{ulem}

\begin{document}

\title{Shape evolution and shape coexistence in Pt isotopes: comparing
  interacting boson  model configuration mixing and Gogny mean-field
  energy surfaces} 
\author{J.E. Garc\'{\i}a-Ramos$^1$, K. Heyde$^2$, L.M.~Robledo$^3$,
  and 
  R. Rodr\'{\i}guez-Guzm\'an$^{4,5}$}  
\affiliation{
$^1$Departamento de F\'{\i}sica Aplicada, Universidad de Huelva,
21071 Huelva, Spain\\
$^2$Department of Physics and Astronomy, Ghent University,
Proeftuinstraat 86, B-9000 Gent, Belgium\\ 
$^2$Departamento de F\'{\i}sica Te\'orica, Universidad Aut\'onoma de Madrid,
28049 Madrid, Spain \\
$^4$Department of Physics and Astronomy, Rice University, Houston,
Texas 77005, USA  \\
$^5$Department of Chemistry, Rice University, Houston, Texas 77005,
USA 
}

\begin{abstract}
The evolution of the total energy surface and the nuclear shape in the isotopic chain
$^{172-194}$Pt are studied in the framework of the interacting boson
model, including configuration mixing. The results are 
compared with a self-consistent Hartree-Fock-Bogoliubov calculation
using the Gogny-D1S interaction
and a good agreement between both approaches shows up. The evolution of the
deformation parameters points towards the presence of two different
coexisting configurations in the region 176 $\leq$ A $\leq$ 186. 
\end{abstract}

\maketitle

%%%%%%%%%%%%%%%%%%%%%%%%%%%%%%%%%%%%%%%%%%%%%%%%%%%%%%%%%%%%%%%%%%% 
%Introduction%%%%%%%%%%%%%%%%%%%%%%%%%%%%%%%%%%%%%%%%%%%%%%%%%%%%%%
%%%%%%%%%%%%%%%%%%%%%%%%%%%%%%%%%%%%%%%%%%%%%%%%%%%%%%%%%%%%%%%%%%% 

\section{Introduction}
\label{sec:intro}

Shape coexistence in atomic nuclei has become a very active field
of research during the last decades and clear signals of its existence
have been obtained at and near proton or neutron closed shells
\cite{heyde11,hey83,wood92}, more in particular in light nuclei with a
closed neutron shell at  N = 8, 20, 28 and 40 closed shells as
well as in heavy nuclei such as the Sn  
%(Z = 50) 
and the Pb nuclei. 
%(Z = 82). 
It seems that, without exception,
shape coexistence is associated with the presence of
low-lying excited $0^+$ states \cite{heyde11}. The Pb region is a very
well-documented example of shape coexistence, both experimental
  and theoretically  (see \cite{heyde11} and references
  therein). Starting at the neutron closed 
shell (at N=126), decreasing in mass until reaching the very neutron-deficient
nuclei, even  
going beyond the mid-shell point at N=104, there is ample experimental evidence
for shape coexisting bands in both the Pb (Z=82) and Hg (Z=80)
nuclei. A question that arises is whether or not this structure, in which
two or more bands coexist, survives when moving away from the closed proton
shell at Z=82. In particular, the Pt nuclei seem to be a good example to test
the survival of the coexisting families of states. It is highly illuminating to
compare the systematics of the low-lying states in the energy spectra 
of the Z=82 proton closed-shell Pb nuclei, the Z=80 Hg nuclei, and the
Z=78 Pt nuclei. Whereas the intruder bands are easily singled out for the Pb and Hg nuclei
in which the excitation energies display the characteristic parabolic pattern as a function
of neutron number N, with a minimal excitation energy around the N=104 neutron mid-shell nucleus, the
intruder structure seems ``lost'' in the Pt nuclei. 
Focusing on the systematics of the energy spectra in the Pt nuclei, 
%as a function of the neutron number,
one observes a rather sudden drop in the excitation energy of the
0$^+_2$, 4$^+_1$,  
2$^+_3$, and 6$^+_1$ states between N=110 
and N=108, followed by a particularly flat behaviour in excitation energy as a function
of neutron number N until the energy of those states start to move up again around neutron 
number N=100.

To study  shape coexistence, there are several approximations
  available. 
Among them we have the 
nuclear shell-model \cite{caurier05} for light nuclei or
the selfconsistent mean-field methods for medium and heavy masses
mostly of the  Hartree-Fock-Bogoliubov (HFB) types
\cite{bender03,Ring80}, as well as their 
beyond mean-field extensions in the spirit of the Generator Coordinate
Method (GCM).
In the nuclear shell-model, shape coexistence is obtained by incorporating many-particle 
many-hole excitations across known closed shells in the model spaces
used, while in selfconsistent mean-field methods 
shape coexistence arises in the form of competing configurations based
on different nuclear shapes labelled by the corresponding intrinsic  
deformations \cite{egido04,rodri10}. We should mention that total
energy surface calculations have also been carried out for heavy
nuclei in the Pb region, starting from a deformed Wood-Saxon
potential \cite{bengt87,bengt89,Naza93}.  
Constrained mean-field
calculations are nowdays routinely perfomed with several effective 
interactions and different levels of sophistication
\cite{duguet03,rodri10}.  
On the other hand, zero point quantum fluctuations 
not explicitly considered at the mean-field level can be
systematically taken into account within the symmetry-projected GCM
\cite{bender03,rodri02}. 
Very recently several works have been carried out in the Pb mass
region either
starting from Skyrme functional
\cite{duguet03,smirnova03,bender04,grahn08,yao13}, using the Gogny
interaction
\cite{girod89,dela94,chasman01,egido04,rodri04,sarri08,rodri10} or the
relativistic mean-field (RMF) approach \cite{sharma92,patra94,yoshida94,yoshida97,fossion06,niksic02,niksic10,
niksic11}.
A third alternative comes from a symmetry dictated truncation of the large
shell-model space, such as the interacting boson model (IBM) \cite{iach87}. 
The IBM starts from the assumption that the low-lying nuclear collective excitations can be described
in terms of bosons with angular momentum
$L=0$ ($s$ bosons) and $L=2$ ($d$ bosons). These building blocks are 
considered to capture the most important nucleon-nucleon correlations in the formation
of nucleon pairs (pairing property and quadrupole collectivity)
corresponding to pairs of nucleons either coupled to angular
momentum $0$ or $2$. The number of valence bosons is counted as 
half the number of valence nucleons, irrespective of their charge
(neutrons or protons) or particle/hole character,  as shown by Otsuka
{\it et al.}~\cite{Otsu78}.

 In the case of the IBM,
shape coexistence arises including two-particle two-hole (2p-2h) (or even higher np-nh) 
excitations across the closed shells, but considering them as extra  
bosons, {\it i.e.}, pairs of nucleons. This extension is called IBM
configuration mixing (IBM-CM for short) \cite{duval82}. An advantage of using the IBM is
the connection with both the shell-model and the mean-field
approach. The extension of the model space in both the nuclear shell-model and the IBM is 
based on enlarging the model space with multi-particle multi-hole excitations. 
A drawback of the IBM results from the fact that an increasing number
of parameters need to be fitted. They are determined by
adjusting to the large body of experimental data, including both
excitation energies and B(E2) reduced transition probabilities, using a
$\chi^2$ fitting procedure. Thereby, the IBM predictive power becomes
curtailed. A possibility to improve the approach is to rely on the
ability of the IBM to derive energy surfaces (mean-field energy)
associated with a given Hamiltonian \cite{Gino80}.
As the mean field parameters are usually adjusted to global
properties like binding energies or nuclear matter properties the range
of applicability of the mean field extends over the whole periodic 
table and makes it a good candidate to be used  
to fit the parameters defining the IBM Hamiltonian. 
It is in this spirit that Nomura {\it et. al.} have recently
explored the connection between the IBM Hamiltonian and the mean-field \cite{nomura08},
mapping the selfconsistent  mean-field energy surfaces onto the IBM space.
In the present paper,  we exploit the  possibility 
of studying the IBM-CM energy surfaces starting from a
Hamiltonian that describes the spectroscopic properties of the chain of isotopes $^{172-194}$Pt. 
This way to proceed is  very different to Nomura's method, because
in that case the Hamiltonian's parameters are fixed from the
mean-field energy surface, while in the present work from the
spectroscopic properties.
   
In two previous papers \cite{Garc09,Garc11}, we used the IBM-CM to
extensively study the Pt nuclei. We carried out a detailed analysis of
the energy spectra and absolute B(E2) values for states up to an
energy $\sim$ $1.5$ MeV. This study allowed us to extract the parameters
describing the IBM-CM Hamiltonian in a precise way and we concluded
that the presence of intruder configurations does not show up
explicitly in the Pt isotopes inspecting the systematics of the
experimental data on energy spectra and B(E2) reduced transition
probabilities (up to an energy of $\sim$ 1.5 MeV), as compared to the
Pb or Hg isotopes. A conclusion was that in the case of the Pt nuclei,
the configuration mixing is somehow ``hidden''.  
Attempts to describe the $^{172-194}$Pt nuclei without 
invoking the effect of intruder excitations have been carried out by
McCutchan {\it et al.}~\cite{Mccu05,Mccu05b,Mccu08}. 
In the present paper
we will use the parameters given in \cite{Garc09,Garc11} without any
change. Our goal is to study the energy surfaces of $^{172-194}$Pt extracted
from phenomenological IBM-CM Hamiltonians, using the intrinsic state 
formalism including configuration mixing and to compare with
selfconsistent mean-field 
calculations, {\it i.e.}, a HFB calculation with a Gogny-D1S 
interaction \cite{rodri10}. 
A successful comparison, even at a qualitative level, will assess the validity
of both viewpoints in the description of shape coexistence.
 
This particular mass region (including the Pt nuclei) 
has been also studied in the framework of the IBM by Nomura {\it
  et al.}~without \cite{nomura11a,nomura11b,nomura11c} and with 
the use of configuration mixing 
\cite{nomura12,nomura13} using as input the total energy 
surfaces derived from HFB calculations (Gogny-D1M interaction), with 
the aim of mapping the mean-field energy surfaces onto IBM energy
surfaces and, therefore, to obtain a set of parameters defining the
IBM Hamiltonians.  
%It is therefore of interest to compare the IBM-CM energy surfaces
%calculated in the present paper with the ones appearing in
%\cite{nomura11a}.

%%%%%%%%%%%%%%%%%%%%%%%%%%%%%%%%%%%%%%%%%%%%%%%%%%%%%%%%%%%%
%IBM-CM %%%%%%%%%%%%%%% 
%%%%%%%%%%%%%%%%%%%%%%%%%%%%%%%%%%%%%%%%%%%%%%%%%%%%%%%%%%%%

\section{The IBM-CM model}
\label{sec:IBM}

The IBM-CM allows the simultaneous treatment and mixing of several
boson configurations which correspond to different particle--hole
(p--h) shell-model excitations \cite{duval82}. In the approach that is
used in the present study no distinction is made between particle- and hole-bosons. 
Thus, the Hamiltonian describing the interacting system of two configurations, one called the
``regular'' configuration, corresponding to N bosons and the other called the ``intruder'' 
configuration, corresponding to N+2 bosons, including the mixing term
between the [N] and [N+2] boson systems, can be written as
\begin{equation} 
  \hat{H}=\hat{P}^{\dag}_{N}\hat{H}^N_{\rm ecqf}\hat{P}_{N}+
  \hat{P}^{\dag}_{N+2}\left(\hat{H}^{N+2}_{\rm ecqf}+
    \Delta^{N+2}\right)\hat{P}_{N+2}\
  +\hat{V}_{\rm mix}^{N,N+2}~,
\label{eq:ibmhamiltonian}
\end{equation}
where $\hat{P}_{N}$ and $\hat{P}_{N+2}$ are projection operators onto
the $[N]$ and the $[N+2]$ boson spaces 
respectively, $\hat{V}_{\rm mix}^{N,N+2}$  describes
the mixing between the $[N]$ and the $[N+2]$ boson subspaces, and
\begin{equation}
  \hat{H}^i_{\rm ecqf}=\varepsilon_i \hat{n}_d+\kappa'_i
  \hat{L}\cdot\hat{L}+
  \kappa_i
  \hat{Q}(\chi_i)\cdot\hat{Q}(\chi_i), \label{eq:cqfhamiltonian}
\end{equation}
is the extended consistent-Q Hamiltonian (ECQF) \cite{warner83}, 
%which is not the most general IBM Hamiltonian,  
with $i=N,N+2$, $\hat{n}_d$ the $d$ boson number operator,
\begin{equation}
  \hat{L}_\mu=[d^\dag\times\tilde{d}]^{(1)}_\mu ,
\label{eq:loperator}
\end{equation}
the angular momentum operator, and
\begin{equation}
  \hat{Q}_\mu(\chi_i)=[s^\dag\times\tilde{d}+ d^\dag\times
  s]^{(2)}_\mu+\chi_i[d^\dag\times\tilde{d}]^{(2)}_\mu~,\label{eq:quadrupoleop}
\end{equation}
the quadrupole operator. 
\begin{figure}[hbt]
\includegraphics[width=0.9\textwidth]{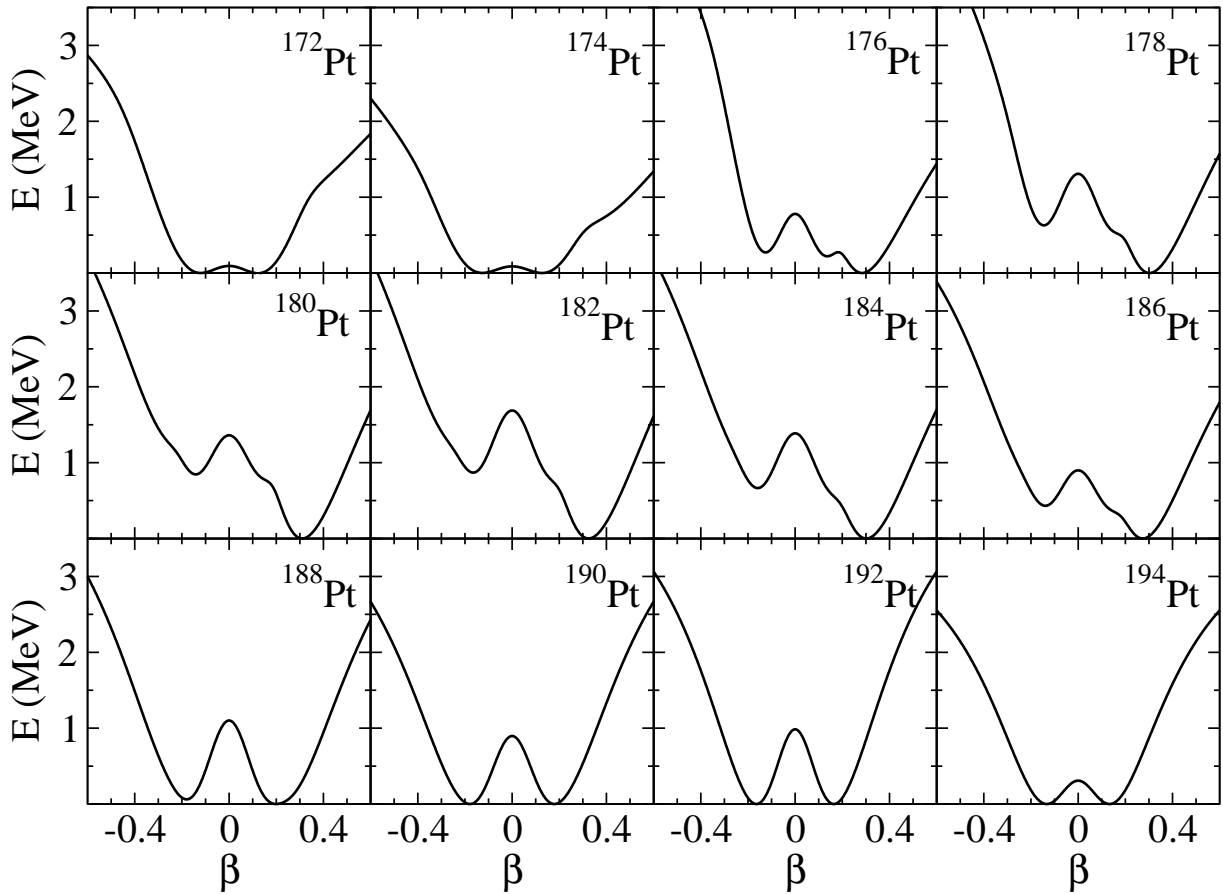}
\caption{IBM-CM total energy curves for $^{172-194}$Pt as a function of
  the $\beta$ deformation parameter (IBM-CM parameters as given in \cite{Garc09}).}
\label{fig-IBM-axial-curves}
\end{figure} 

The parameter $\Delta^{N+2}$ can be
associated with the energy needed to excite two particles across the
$Z=82$ shell gap, corrected for the pairing interaction energy gain and including
monopole effects~\cite{hey85,hey87}.
The operator $\hat{V}_{\rm mix}^{N,N+2}$ describes the mixing between
the $[N]$ and the $[N+2]$ configurations and is defined as
\begin{equation}
  \hat{V}_{\rm mix}^{N,N+2}=w_0^{N,N+2}(s^\dag\times s^\dag + s\times
  s)+w_2^{N,N+2} (d^\dag\times d^\dag+\tilde{d}\times \tilde{d})^{(0)}.
\label{eq:vmix}
\end{equation}

The considered Hamiltonian is not the most general one in each Hilbert
space, [N] and [N+2], but this
approach has been shown to be a rather good approximation in many
realistic calculations \cite{iach87}.
In particular, in the present study, we have taken the parameters
obtained in \cite{Garc09}. 

These parameters are quite different from the ones obtained by
Nomura {\it et al.}~in \cite{nomura11a} mostly due to the following reasons:
First, the present IBM-CM makes no distinction  between proton and
neutron bosons, while \cite{nomura11a} do
not use configuration mixing but take into account the distinction between
protons and neutrons, {\it i.e.}, they use IBM-2
\cite{iach87}. Second, we obtain the Hamiltonian's parameters directly
from experimental spectroscopic properties, while in \cite{nomura11a}
the authors extract them from a selfconsistent mean-field energy
surface, through a mapping procedure.

%%%%%%%%%%%%%%%%%%%%%%%%%%%%%%%%%%%%%%%%%%%%%%%%%%%%%%%%%%%%%%%%%%%%%%%%%%%%%%
%Geometry IBM-CM%%%% 
%%%%%%%%%%%%%%%%%%%%%%%%%%%%%%%%%%%%%%%%%%%%%%%%%%%%%%%%%%%%%%%%%%%%%%%%%%%%%%
\begin{figure}[hbt]
\includegraphics[width=0.9\textwidth]{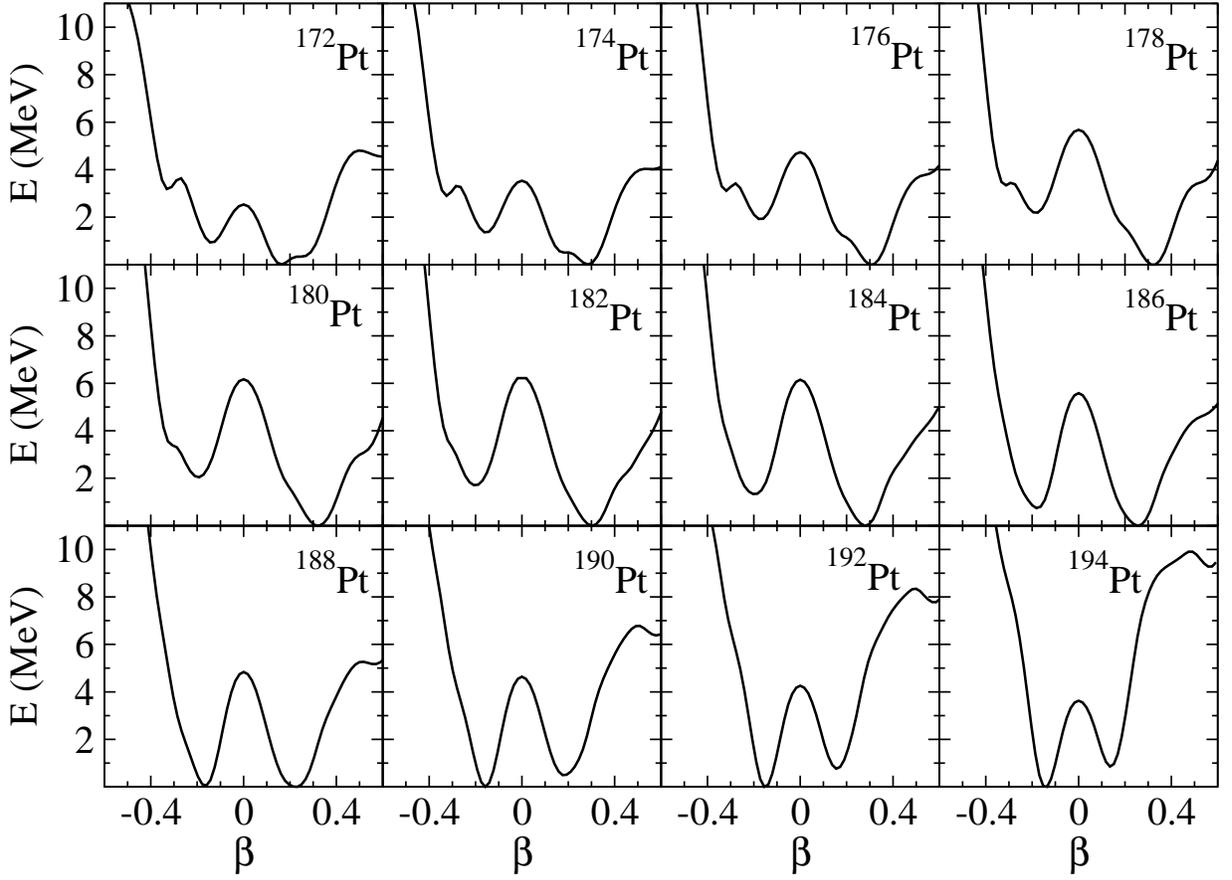}
\caption{ Hartree-Fock-Bogoliubov total energy curves for
  $^{172-194}$Pt as a function of 
  the axial quadrupole deformation parameter $\beta$ (see also
  Ref.~\cite{rodri10}).}  
\label{fig-HFB-axial-curves}
\end{figure} 

\section{Energy surfaces}
\label{sec:energy}

In the eighty's a geometric interpretation of the IBM  was proposed by
Ginocchio and Kirson \cite{Gino80}, using the so-called intrinsic state formalism.  
To define the intrinsic state, one
assumes that the dynamical behavior of the system can be described using
independent bosons moving in an average field. The ground state of the system is a condensate
$|N; \beta_B,\gamma_B \rangle$ of N bosons (the $B$ subindex stands for
boson, as used in Ref.~\cite{nomura08}), occupying the lowest-energy
phonon state, 
$\Gamma^\dag_c$,  
\begin{equation}
\label{GS}
|N; \beta_B,\gamma_B   \rangle = {1 \over \sqrt{N!}} (\Gamma^\dagger_c)^N | 0 \rangle,
\end{equation}
where
\begin{equation}
\label{bc}
\Gamma^\dagger_c = {1 \over \sqrt{1+\beta_B^2}} \left (s^\dagger + \beta_B
\cos     \gamma_B          \,d^\dagger_0          +{1\over\sqrt{2}}\beta_B
\sin\gamma_B\,(d^\dagger_2+d^\dagger_{-2}) \right) ,
\end{equation}
$d^\dagger_\mu$ corresponds to the $\mu$ component of the $d^\dagger$
operator, $\beta_B$ and $\gamma_B$ are variational parameters related with the shape
variables in the geometrical collective model ~\cite{Bohr75}, and the reference boson vacuum state, containing no
s nor d bosons is denoted by the ket vector $|0\rangle$. 
The expectation value
of the Hamiltonian in the intrinsic state (\ref{GS}) provides  
the energy surface of the system, $E(N,\beta_B,\gamma_B)=\langle N; \beta_B,\gamma_B 
|\hat H| N; \beta_B,\gamma_B  \rangle$. The values of $\beta_B$ 
and $\gamma_B$ which minimize the expectation value of the energy, {\it i.e.}, 
the equilibrium values, provide the shape of the nucleus. 

The extension of the IBM geometrical picture in order to include configuration mixing was
proposed by Frank {\it et al.}~\cite{Frank02,Frank04,Frank06,Mora08}
by means of a matrix coherent-state method that allows to describe shape coexistence phenomena.   
The way to proceed is to define a model space with 
states $|N; \beta_B,\gamma_B  \rangle$, $|\ N +2; \beta_B,\gamma_B  \rangle$ 
and to diagonalize the Hamiltonian (\ref{eq:ibmhamiltonian}). Therefore, one needs to
construct the $2\times 2$ matrix:
\begin{equation}
H_{CM}=\left (
\begin{array}{cc}
E(N,\beta_B,\gamma_B)& \Omega(\beta_B)\\
\Omega(\beta_B)& E(N+2,\beta_B,\gamma_B)
\end{array}
\right ) .
\label{surf-cm}
\end{equation}
The diagonal terms $E(N,\beta_B,\gamma_B)$ and $E(N+2,\beta_B,\gamma_B)$ only contain
the $N$ and the $N+2$ contributions of the Hamiltonian (\ref{eq:ibmhamiltonian}),
respectively, while $\Omega(\beta_B)$ corresponds to the 
matrix element describing the mixing of the $[N]$ and $[N+2]$ boson configurations.
The expressions for both the diagonal and non-diagonal matrix elements (see Ref.~\cite{Mora08}) are: 
\begin{eqnarray}
\nonumber
E_i(N_i,\beta_B,\gamma_B)&=&(\varepsilon_i+6\kappa'_i)\frac{N_i\beta_B^2}{1+\beta_B^2}+\kappa_i
\left(
\frac{N_i}{1+\beta_B^2}(5+(1+\chi_i^2)\beta_B^2)+\frac{N_i(N_i-1)}{(1+\beta_B^2)^2}\right
.\\
&\times&\left.\big(\frac{2}{7}\chi_i^2\beta_B^4-4\sqrt{\frac{2}{7}}
\chi_i\beta_B^3\cos{3\gamma_B}+4\beta_B^2\big)\right),\\
\Omega(\beta_B)&=&\frac{\sqrt{(N_i+2)(N_i+1)}}{1+\beta_B^2}\left
    (w_0^{N,N+2} +w_2^{N,N+2}\frac{\beta_B^2}{\sqrt{5}}\right). 
\end{eqnarray} 
To obtain the energy surface one has to
diagonalize (\ref{surf-cm}) and to consider the lowest eigenvalue.
\begin{figure}
\includegraphics[width=.5\textwidth]{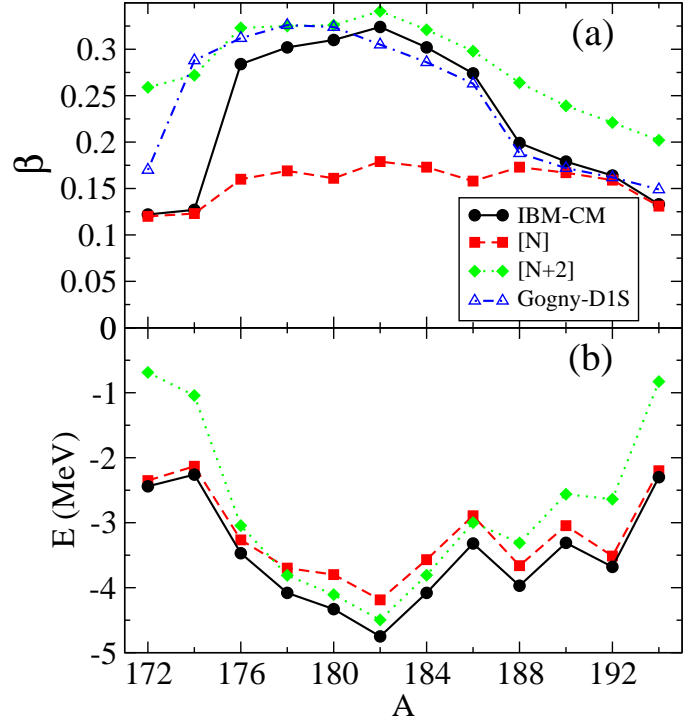}%
\caption{(Color online) (a): Comparison of the value of $\beta$  for IBM-CM and for
  selfconsistent HFB. 
  We also plot the value of $\beta$ corresponding with the 
  unperturbed ``regular'', $[N]$,  and the unperturbed ``intruder'', $[N+2]$,
  boson configurations. (b): energy of the unperturbed regular and intruder IBM 
  bandheads, compared with the ground state energy (full IBM-CM diagonalization).} 
\label{fig-comp-beta}
\end{figure} 

%%%%%%%%%%%%%%%%%%%%%%%%%%%%%%%%%%%%%%%%%%%%%%%%%%%%%%%%%%%%%%
% HFB %%%%%%%%%%%%%%%%%%%%%%%%%%%%%%%%%%%%%%%%%%%%%%%%%%%%%%%%
%%%%%%%%%%%%%%%%%%%%%%%%%%%%%%%%%%%%%%%%%%%%%%%%%%%%%%%%%%%%%%

In the present study, with the aim to obtain the corresponding 
mean-field surfaces for the considered nuclei, we have resorted to
the HFB \cite{Ring80} approximation based on the 
parametrization D1S of the Gogny interaction \cite{rodri10}.
In the calculation of the total energy surfaces for the Pt nuclei
\cite{rodri10}, the HFB
quasiparticle operators have been expanded in a harmonic oscillator basis
containing enough shells (N$_{shell}$=13 major shells in the present case) in order to guarantee
convergence for all values of the mass quadrupole operator and for
all the Pt isotopes. In order to construct the total energy contour plots in the 
$(\beta-\gamma)$ plane, we have to constrain the different components of
the quadrupole operator:
\begin{equation}
Q_{20}=\frac{1}{2} \langle \Phi_{HFB}| 2z^2-x^2-y^2|\Phi_{HFB}\rangle,
\end{equation}
 
\begin{equation}
Q_{22}=\frac{\sqrt{3}}{2} \langle \Phi_{HFB}| x^2-y^2|\Phi_{HFB}\rangle,
\end{equation}

\begin{equation}
Q=\sqrt{Q_{20}^2+Q_{22}^2},
\end{equation}

\begin{equation}
\tan\gamma=\frac{Q_{22}}{Q_{20}}.
\end{equation}
The relationship between $Q$ and $\beta$ can be found in \cite{Bohr75}
and results to be,
\begin{equation}
\beta=\sqrt{\frac{4\pi}{5}}\frac{Q}{A\langle r^2\rangle},
\end{equation} 
where $\langle r^2\rangle=3/5 r_0^2 A^{2/3}$ ($r_0=1.2$ fm). Further
details can be found in \cite{Robl09}.

To compare IBM-CM and HFB total energy surfaces one needs to
establish a relationship between ($\beta_B,\gamma_B$) and ($\beta,\gamma$). 
This problem was first studied 
by Ginocchio and Kirson \cite{Gino80} with as a result that $\gamma_B=\gamma$ and  
the semi-quantitative relationship,
\begin{equation}
\beta\leq 1.18 \frac{2N}{A} \beta_B.
\label{beta}
\end{equation} 
The latter provides the functional dependence between
$\beta$ and $\beta_B$ and establishes that $\beta$ is much smaller
than $\beta_B$, however the precise mapping is an involved
task.  
Moreover, expression (\ref{beta}) should be modified because of the
presence of two different configurations with different number of
active bosons, $N$ and $N+2$, respectively. Extending Eq.~(\ref{beta}) to
the IBM-CM approach, one should sum up the contributions from both configurations
weighted with the square of the amplitude of the wave function in the
$N$ space, $\omega$ (see \cite{Garc09, Garc11}),
\begin{equation}
\beta=
%1.18 \frac{2}{A} \beta_B (N\times\omega+(N+2)\times(1-\omega))= 
1.18 \frac{2}{A} \beta_B (N+2~(1-\omega)).
\label{beta-cm}
\end{equation} 
Comparing the value of $\beta$ resulting from both approaches one 
cannot expect an exact agreement (according to \cite{Gino80} ``we should
not take the actual numerical values (of $\beta$) too seriously''). 
To improve the results we will use as an {\it ansatz},
\begin{equation}
\beta= 1.18 \frac{2}{A} \beta_B (N+2~(1-\omega)) ~ \delta +\xi.
\label{beta2}
\end{equation} 
We have performed a least-squares fit between the IBM and HFB $\beta$
equilibrium values, obtaining $\delta=1.37$ and $\xi=0.07$. In section
\ref{sec:results} all the results presented for the IBM-CM make use
of the above scale transformation (\ref{beta2}).
In \cite{nomura08}, the authors established the connection between
$\beta_B$ and $\beta$ through a mapping procedure among the mean-field
and the IBM energy surfaces.  

%%%%%%%%%%%%%%%%%%%%%%%%%%%%%%%%%%%%%%%%%%%%%%%%%%%%%%%%%%%%%%%%%
%Comparison%%%%%%%%%%%%%%%%%%%%%%%%
%%%%%%%%%%%%%%%%%%%%%%%%%%%%%%%%%%%%%%%%%%%%%%%%%%%%%%%%%%%%%%%%%

\section{Results and discussion}
\label{sec:results}

In Fig.~\ref{fig-IBM-axial-curves}, we show for
every Pt isotope (covering the mass interval 172 $\leq$ A $\leq$ 194) the IBM-CM energy curves along
the axial symmetry axis, as a function of the deformation parameter,
$\beta$. Prolate shapes correspond to $\beta>0$ while oblate shapes to
$\beta<0$. These curves correspond to the 
lowest eigenvalue of the matrix (\ref{surf-cm}). The lightest isotopes,
$^{172-174}$Pt, present two very shallow degenerate minima, oblate
and prolate, respectively, that correspond to a small value of $\beta$,
%$\beta\approx 0.12$ 
with a rather modest
barrier at $\beta=0$. Consequently, in the present approach, these isotopes are
close to exhibiting a spherical shape. 
The next isotope, $^{176}$Pt, starts to develop a more pronounced
deformed minimum. Indeed two prolate and an oblate minimum 
%(see Fig.~\ref{fig_ibm_ener_surph}) 
are observed for this isotope,
the deeper minimum corresponding to the more deformed minimum with
$\beta\approx 0.3$.
The $^{178-186}$Pt isotopes show a similar structure, with
a well-deformed prolate minimum, $\beta\approx 0.3$, a large barrier at
$\beta=0$, of about $1.5$ MeV, 
and an oblate local minimum. Finally, the $^{188-194}$Pt isotopes exhibit two
quasi-degenerate minima, at the same value of $\beta$, the first prolate
while the second oblate, with an equilibrium value of $\beta$
which decreases from $\beta\approx 0.2$ for $^{188}$Pt towards
$\beta\approx 0.13$ in $^{194}$Pt.  
The height of the central barrier remains almost constant for
$^{188-192}$Pt at a  value $\approx$ 1 MeV, but is strongly reduced
to $0.3$ MeV in $^{194}$Pt. 
\begin{figure*}
\includegraphics[width=0.25\textwidth]{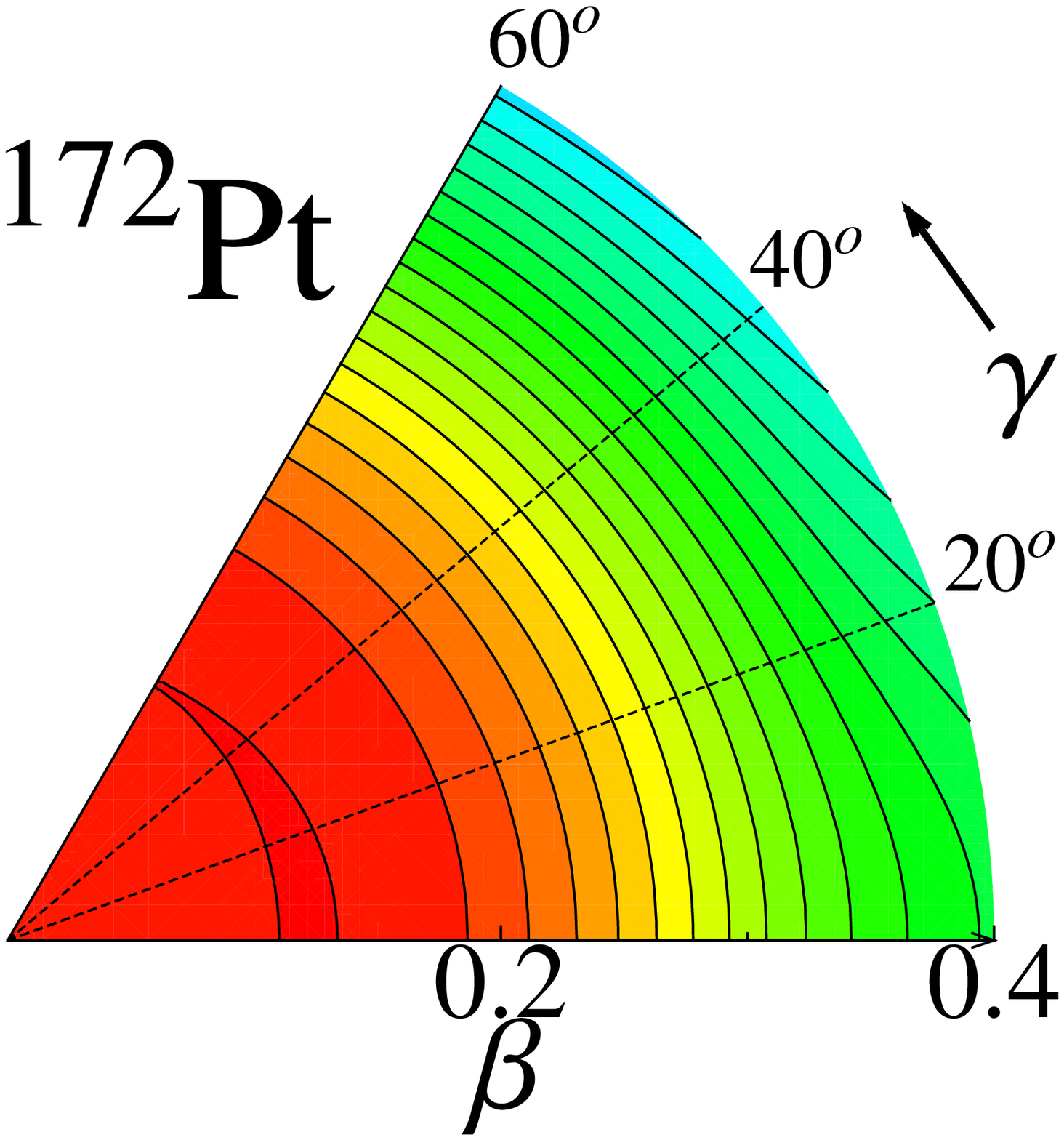}%
\includegraphics[width=0.25\textwidth]{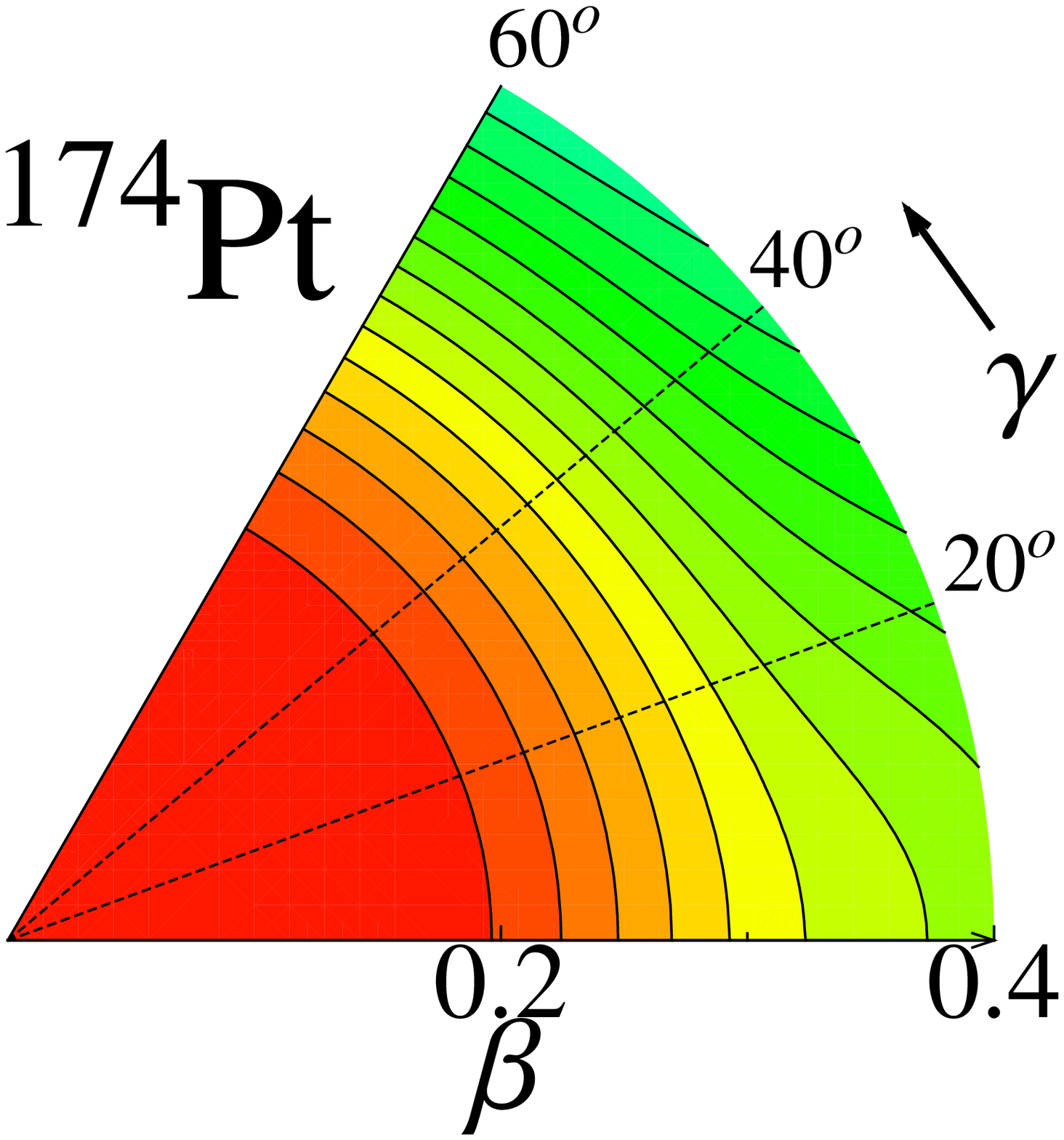}%
\includegraphics[width=0.25\textwidth]{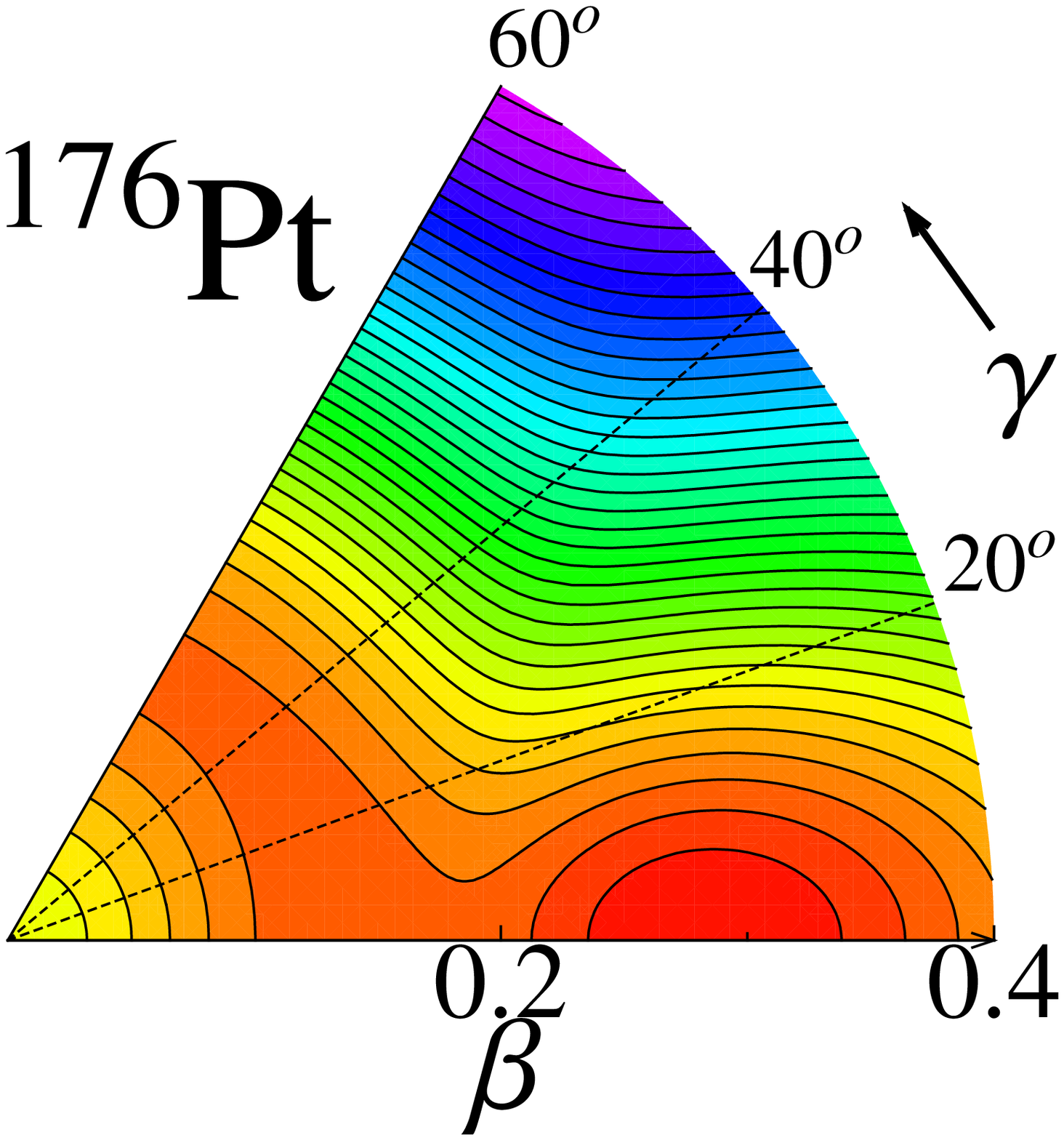}%
\includegraphics[width=0.25\textwidth]{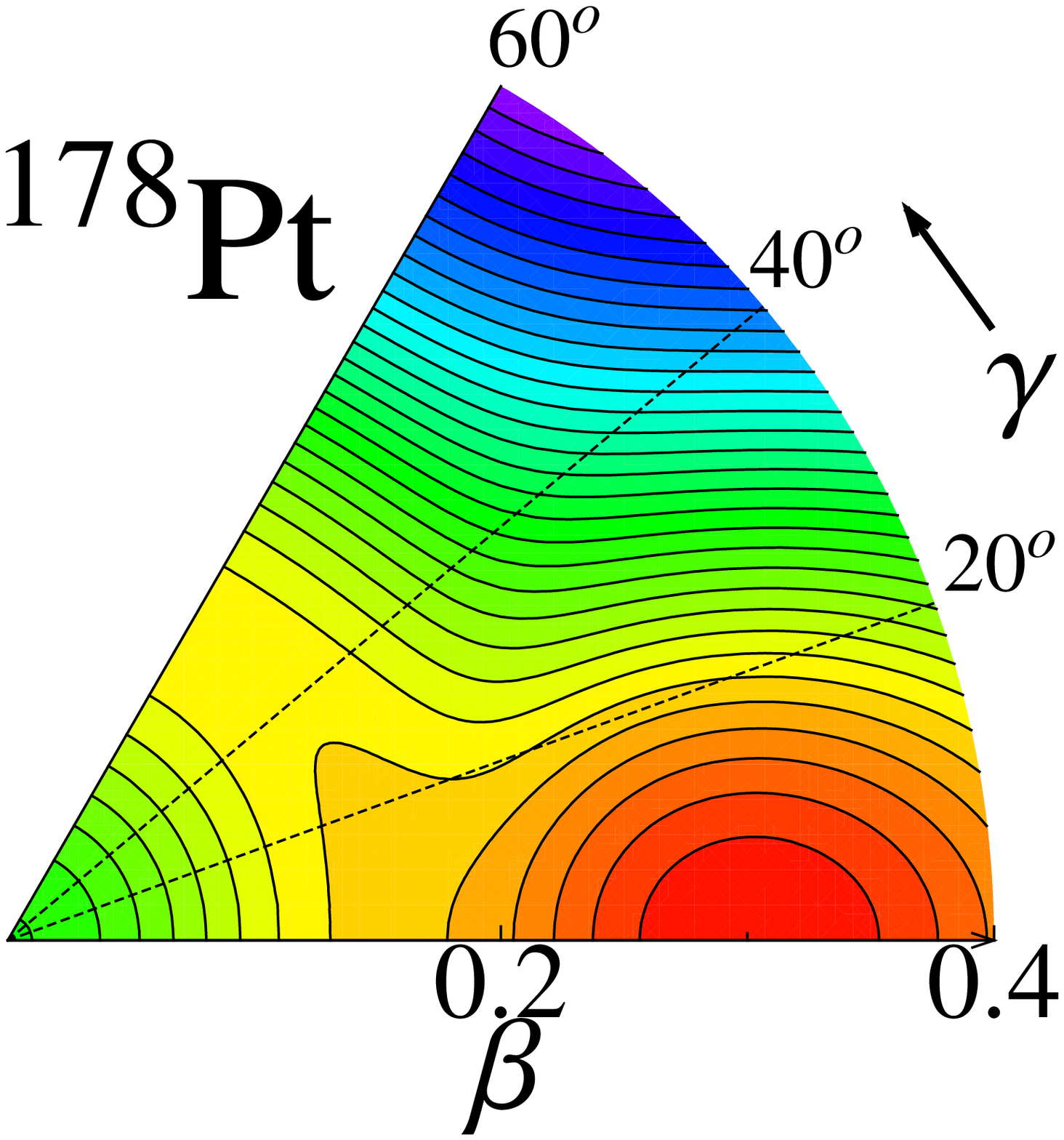}\par
\includegraphics[width=0.25\textwidth]{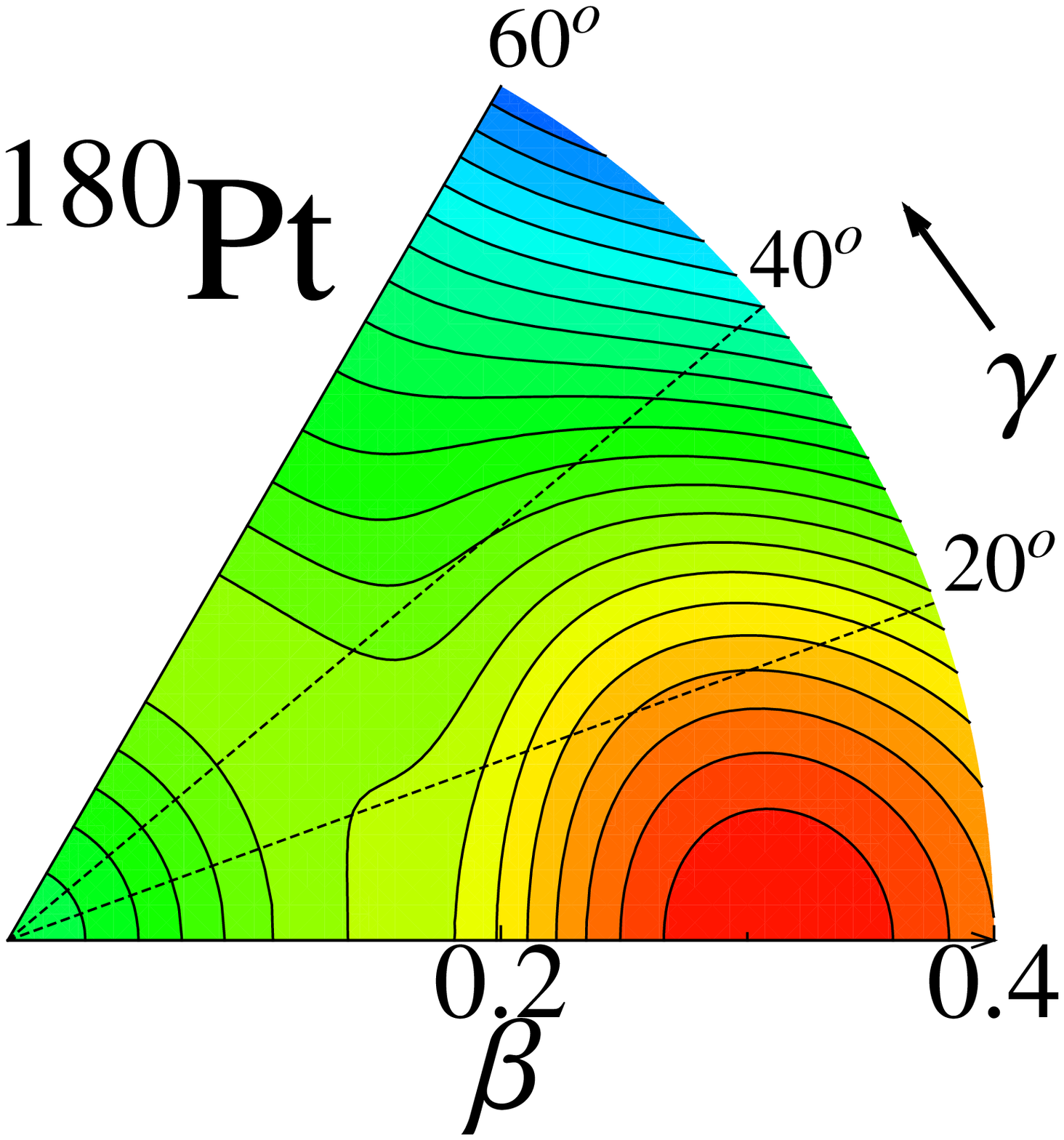}%
\includegraphics[width=0.25\textwidth]{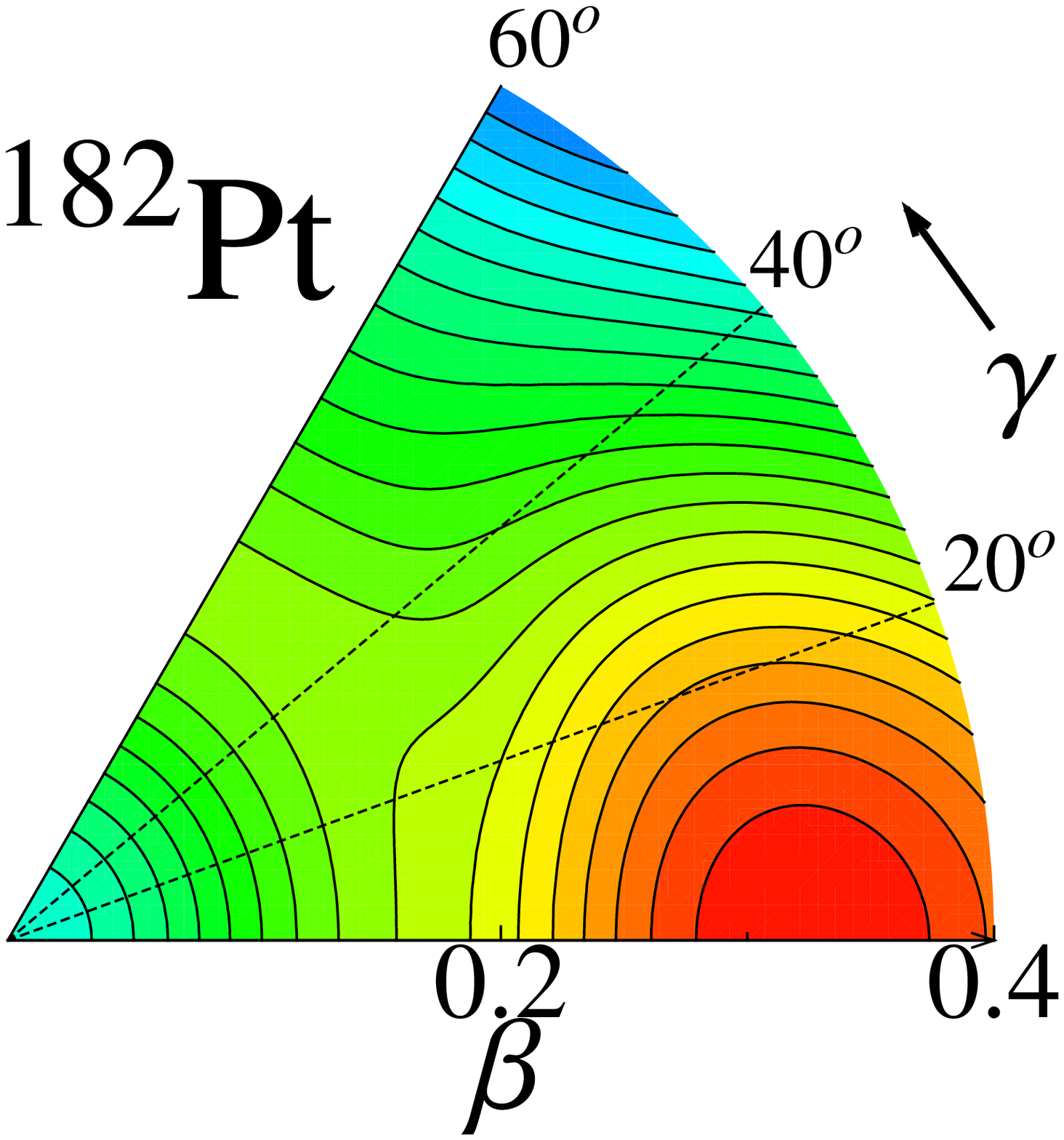}%
\includegraphics[width=0.25\textwidth]{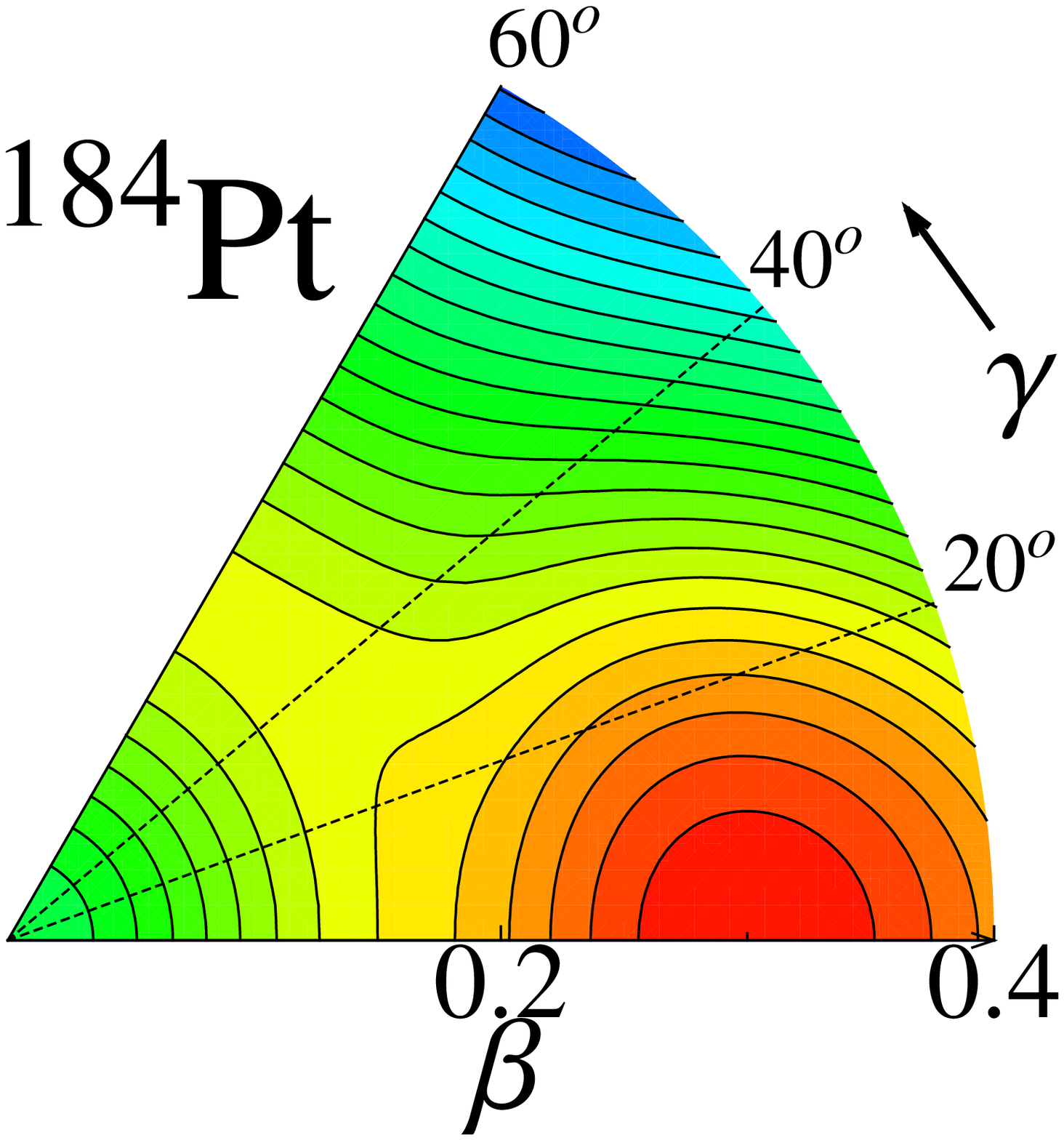}%
\includegraphics[width=0.25\textwidth]{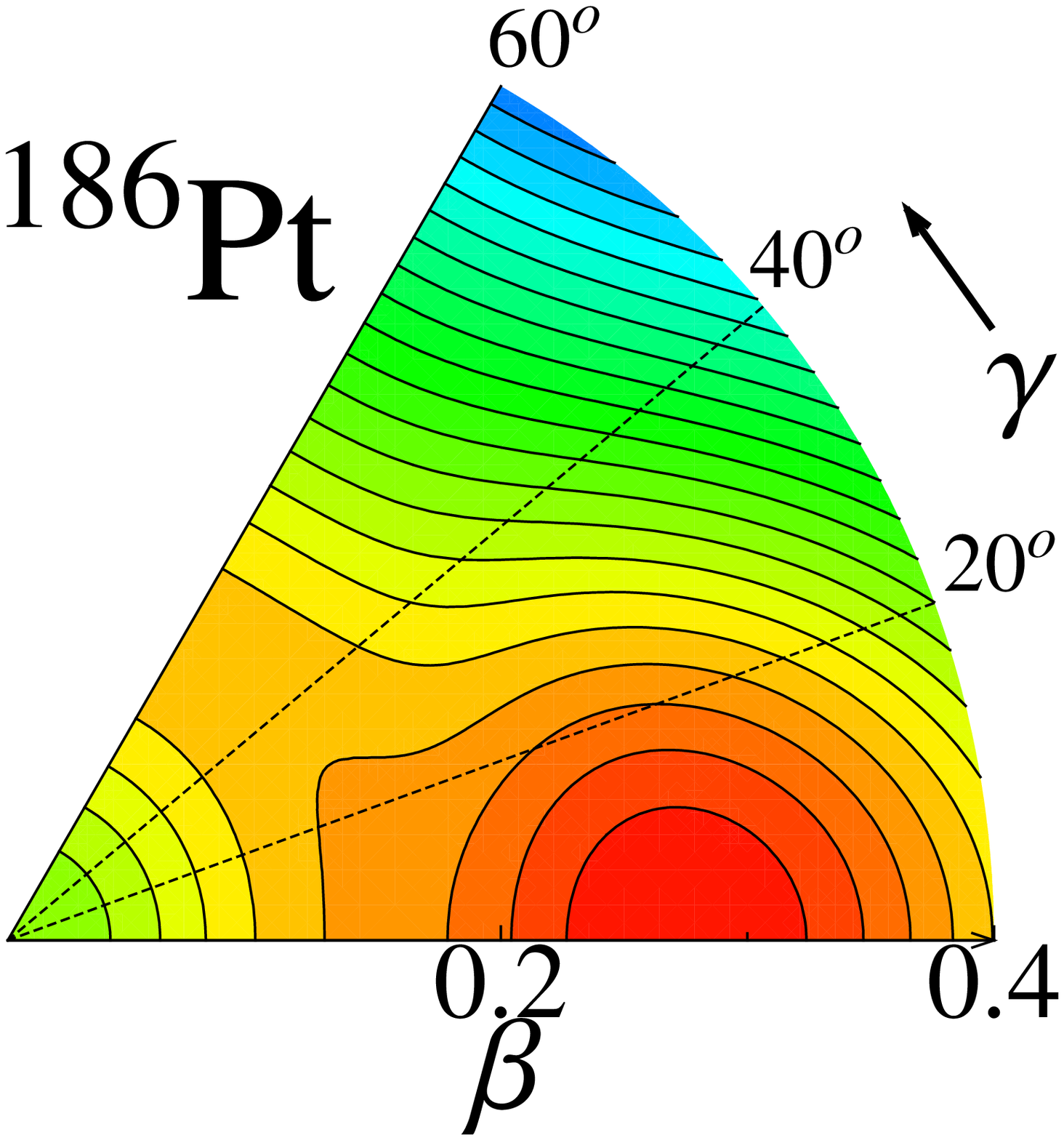}\par
\includegraphics[width=0.25\textwidth]{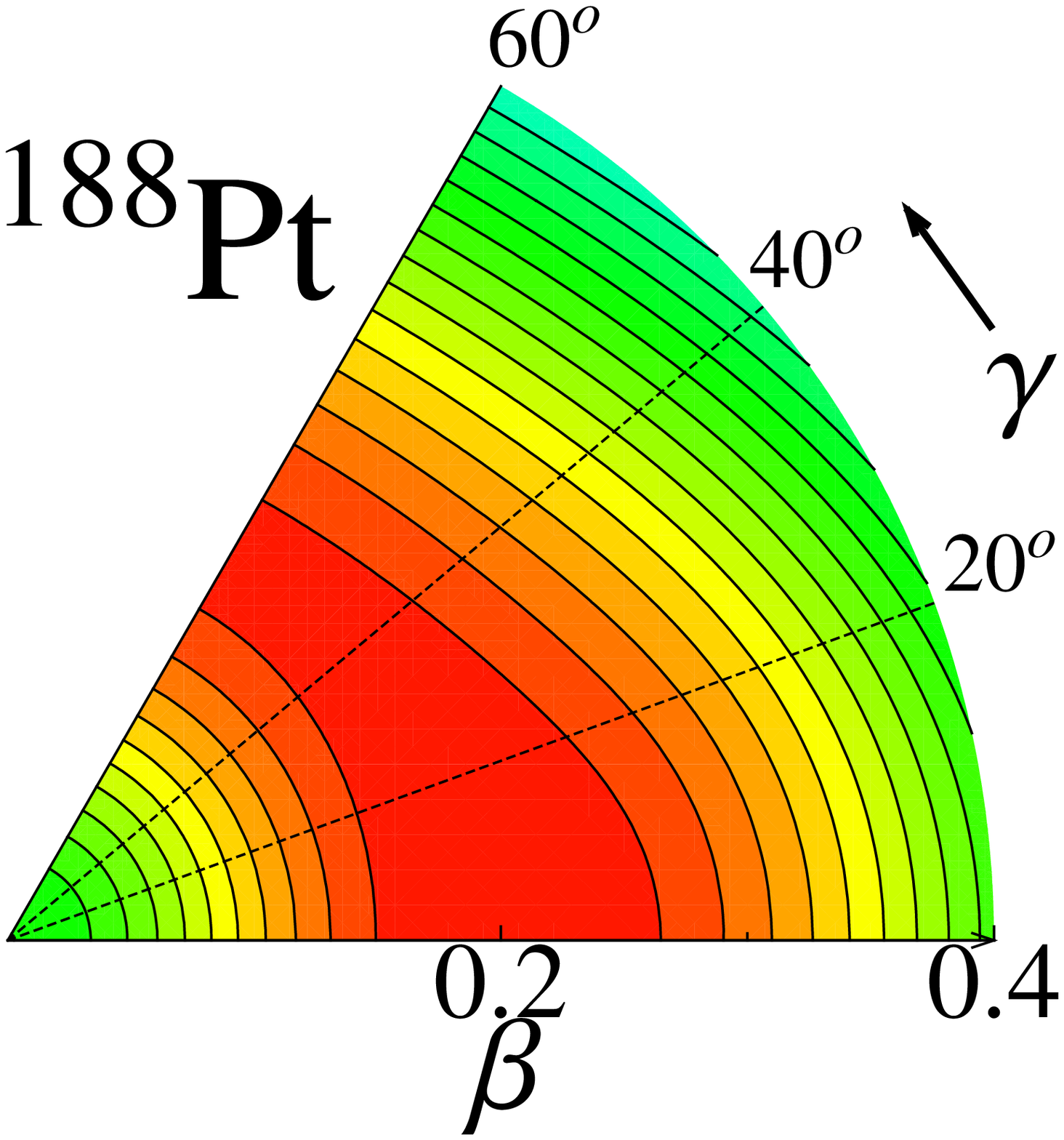}%
\includegraphics[width=0.25\textwidth]{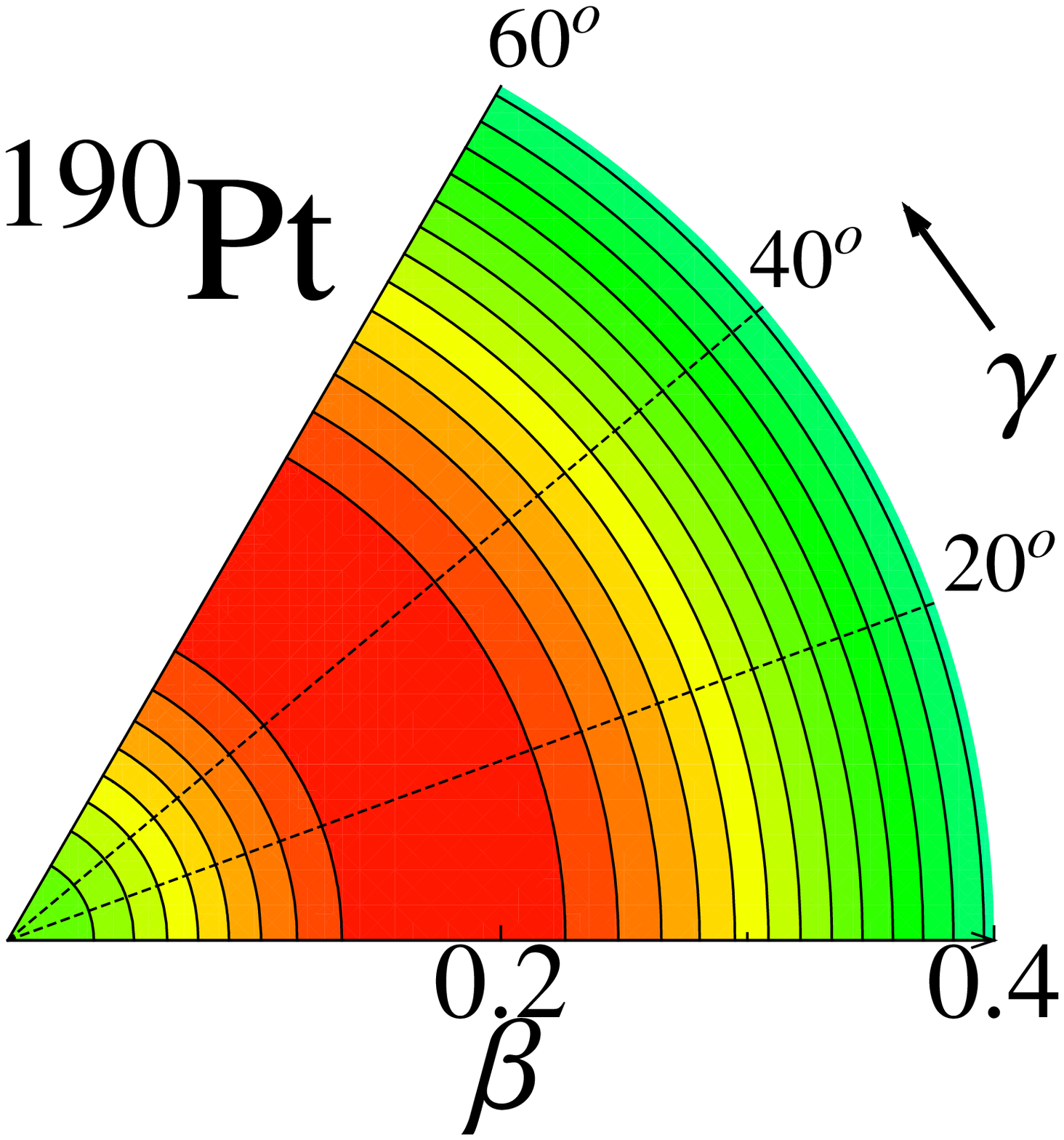}%
\includegraphics[width=0.25\textwidth]{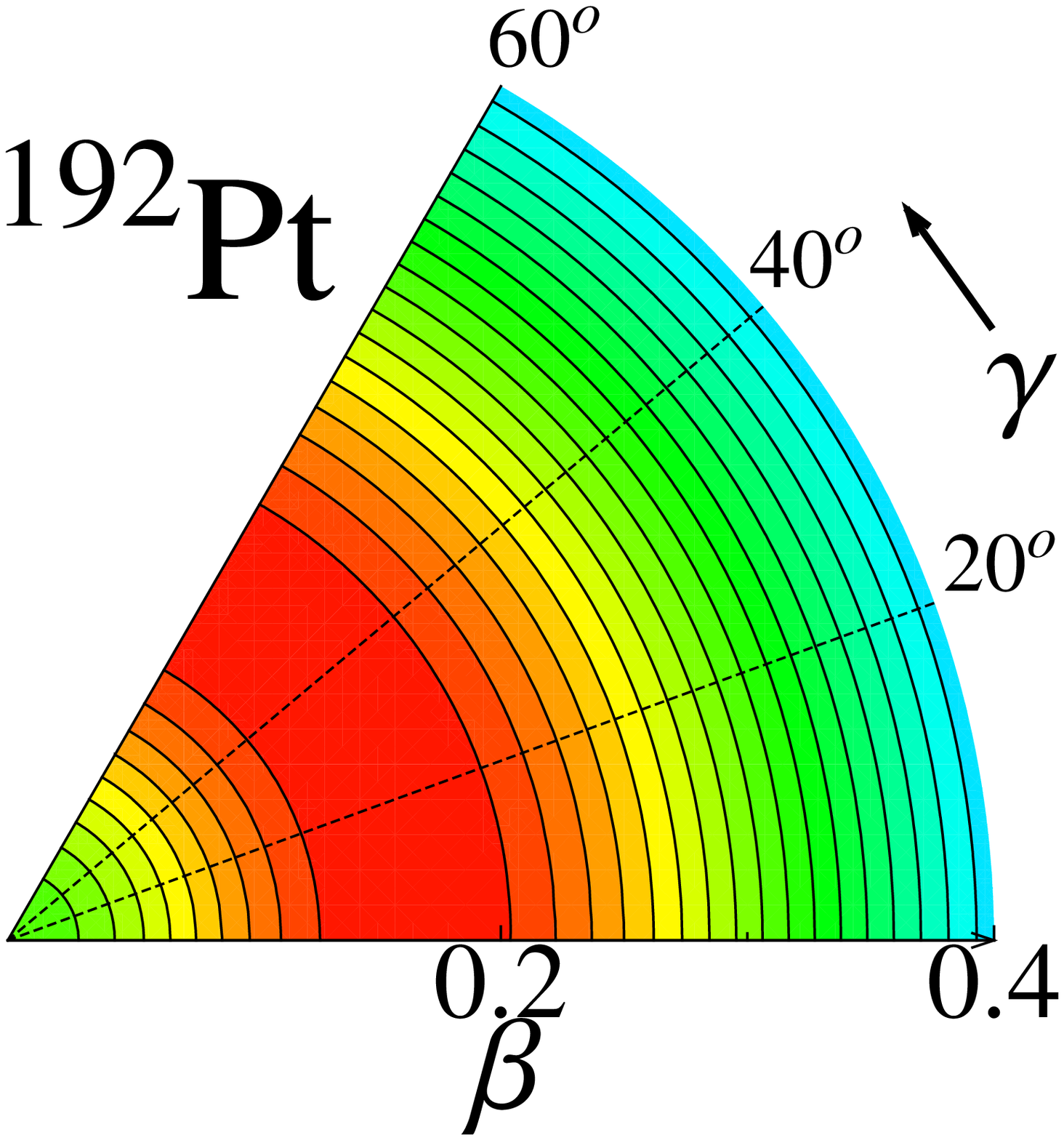}%
\includegraphics[width=0.25\textwidth]{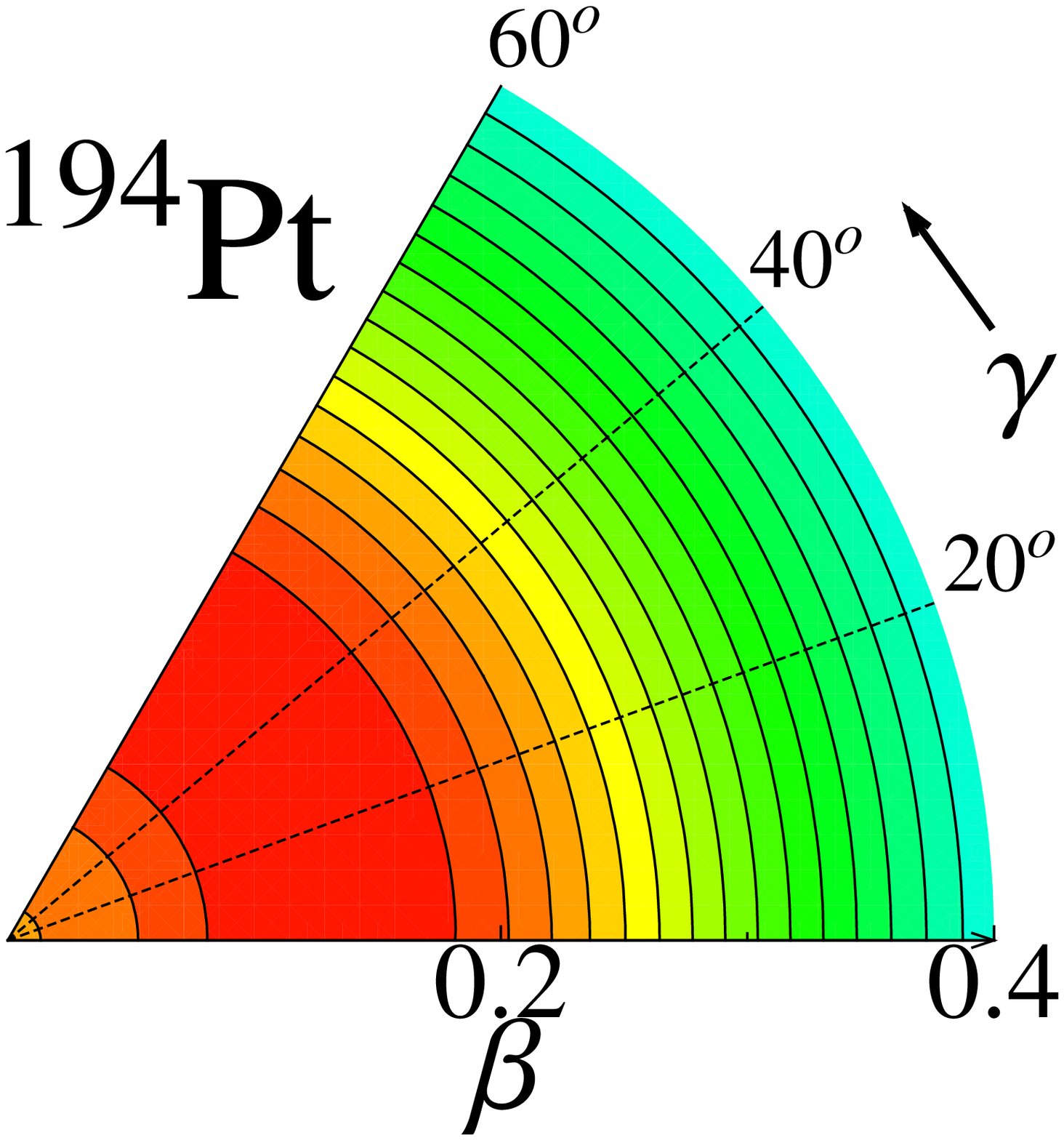}
\caption{(Color online) IBM-CM contour plots for $^{172-194}$Pt as a function of
  $\beta$ and $\gamma$ (IBM-CM parameters as given in \cite{Garc09}).
    The separation between adjacent contour lines amounts to $100$ keV.
    The deepest energy minimum is set to zero, corresponding to
    the red color, while green corresponds to $\approx 3$ MeV.}
\label{fig_ibm_ener_surph}
\end{figure*} 

\begin{figure*}
\includegraphics[width=0.25\textwidth]{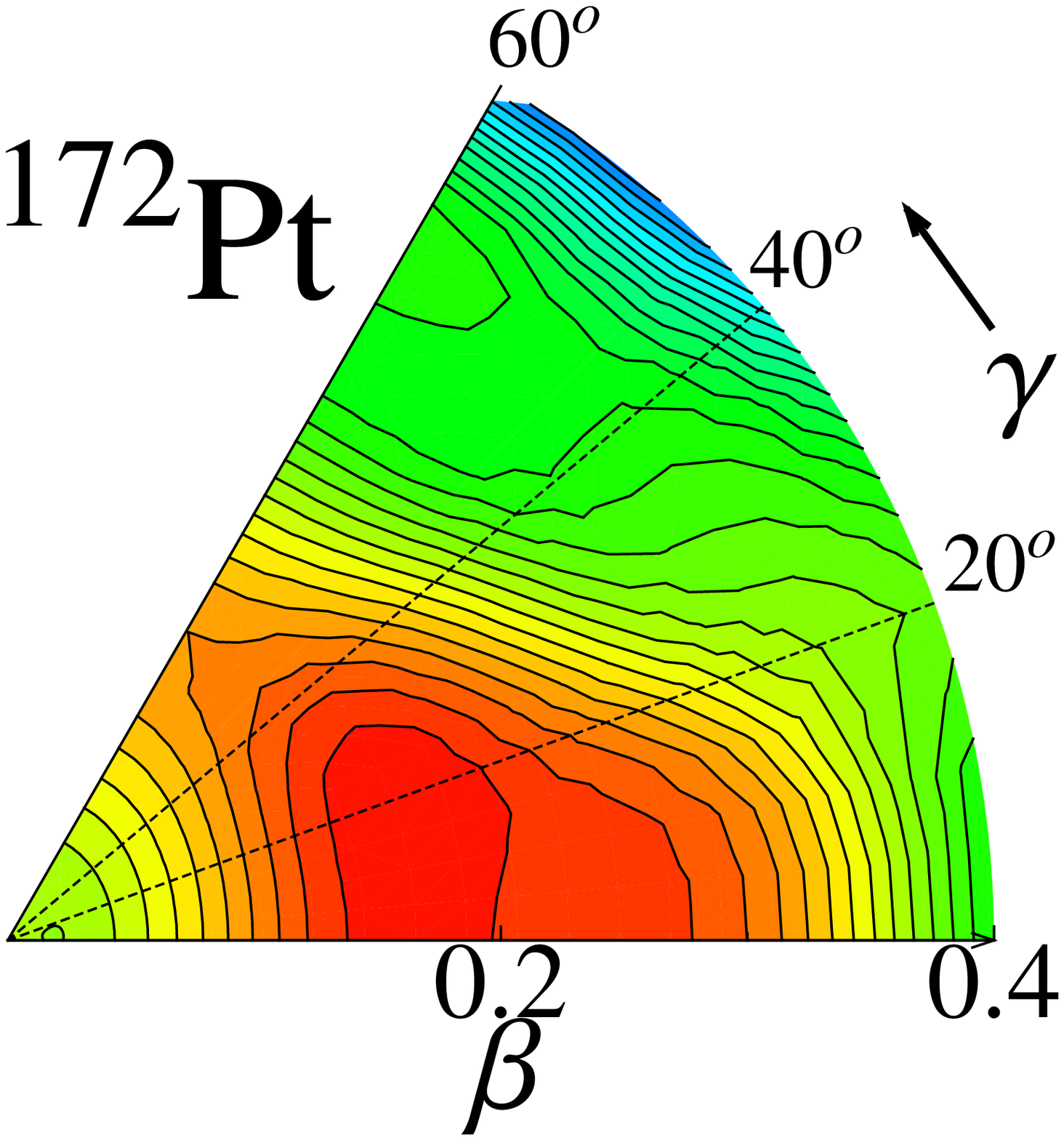}%
\includegraphics[width=0.25\textwidth]{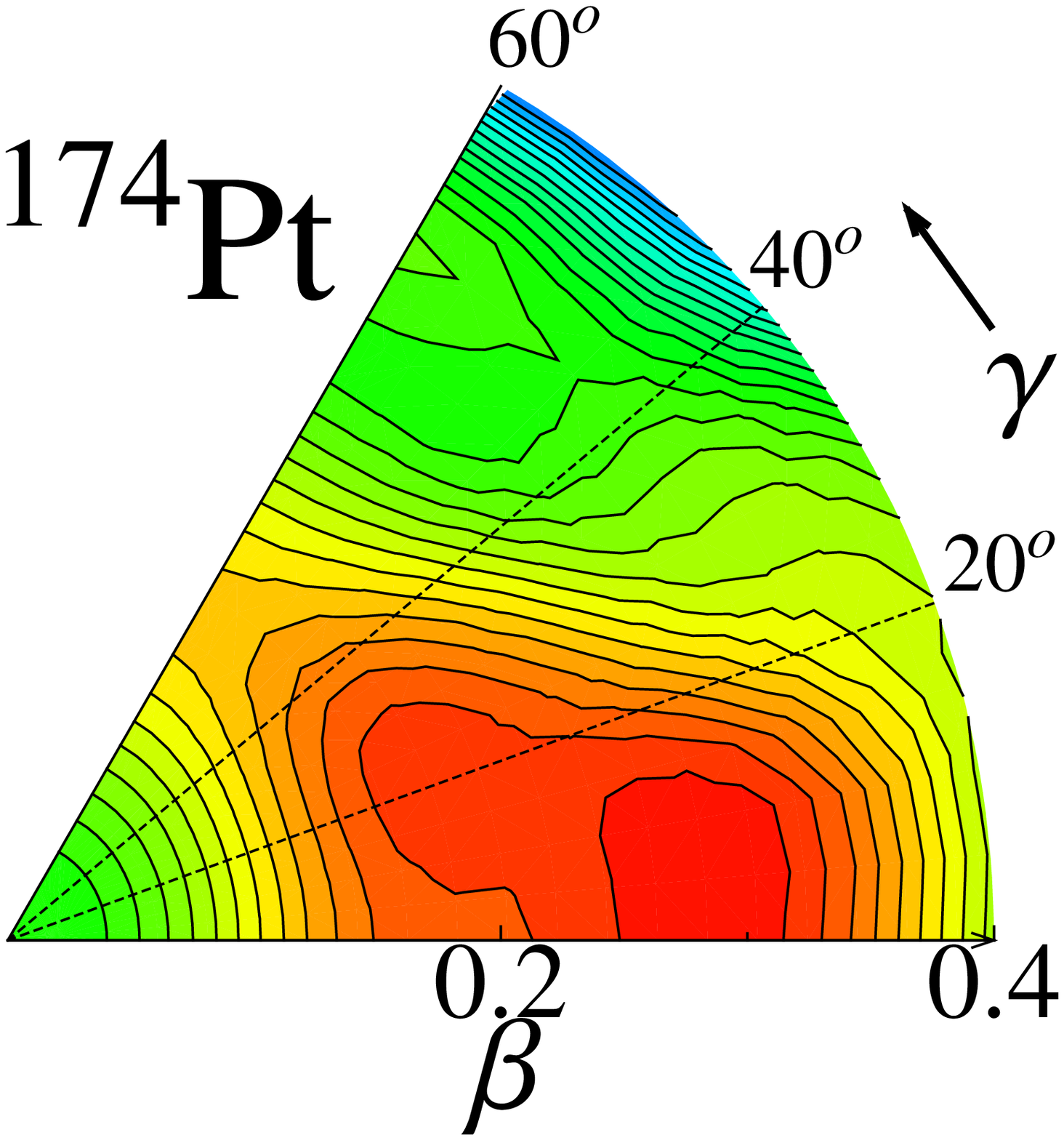}%
\includegraphics[width=0.25\textwidth]{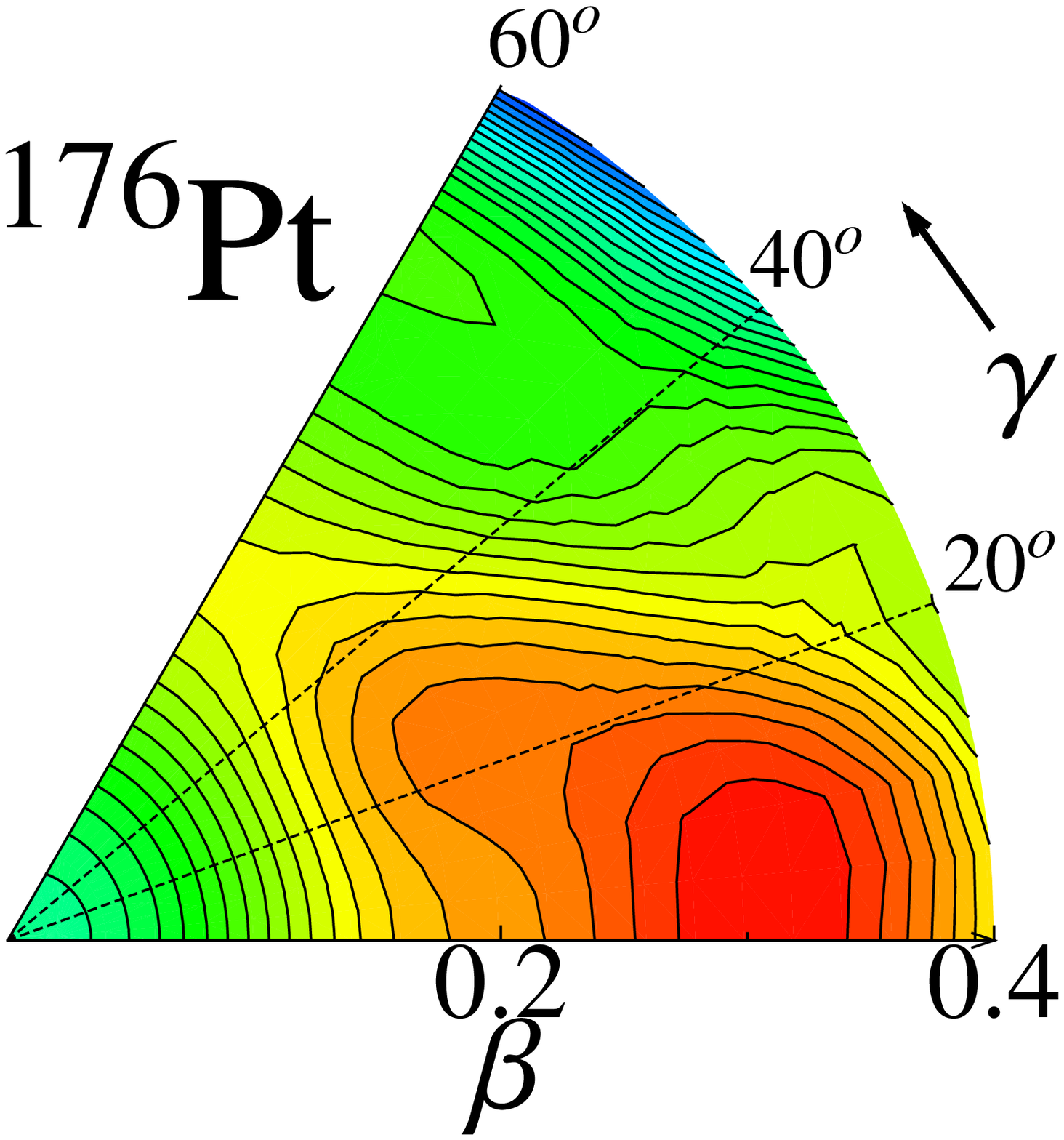}%
\includegraphics[width=0.25\textwidth]{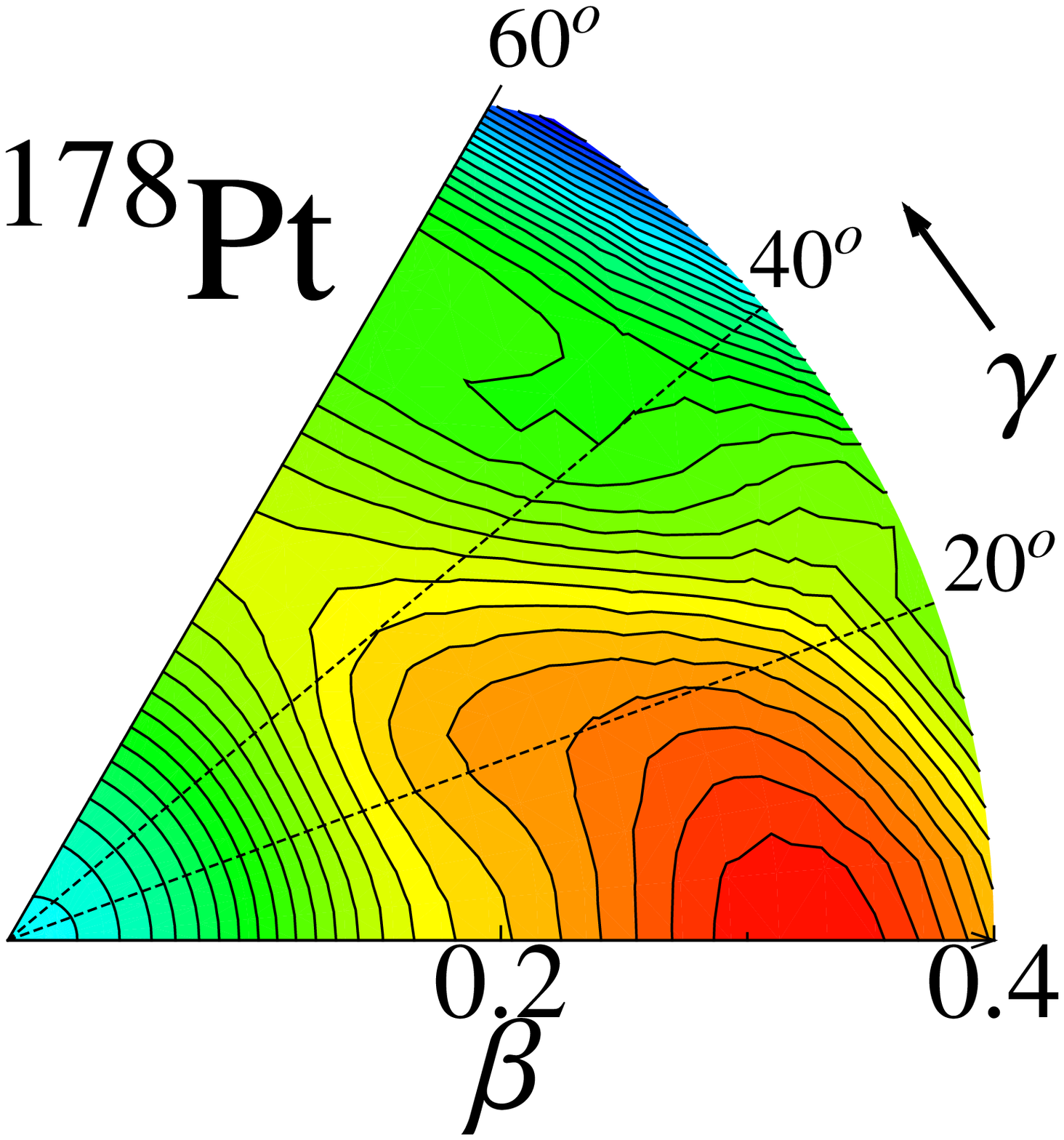}\par
\includegraphics[width=0.25\textwidth]{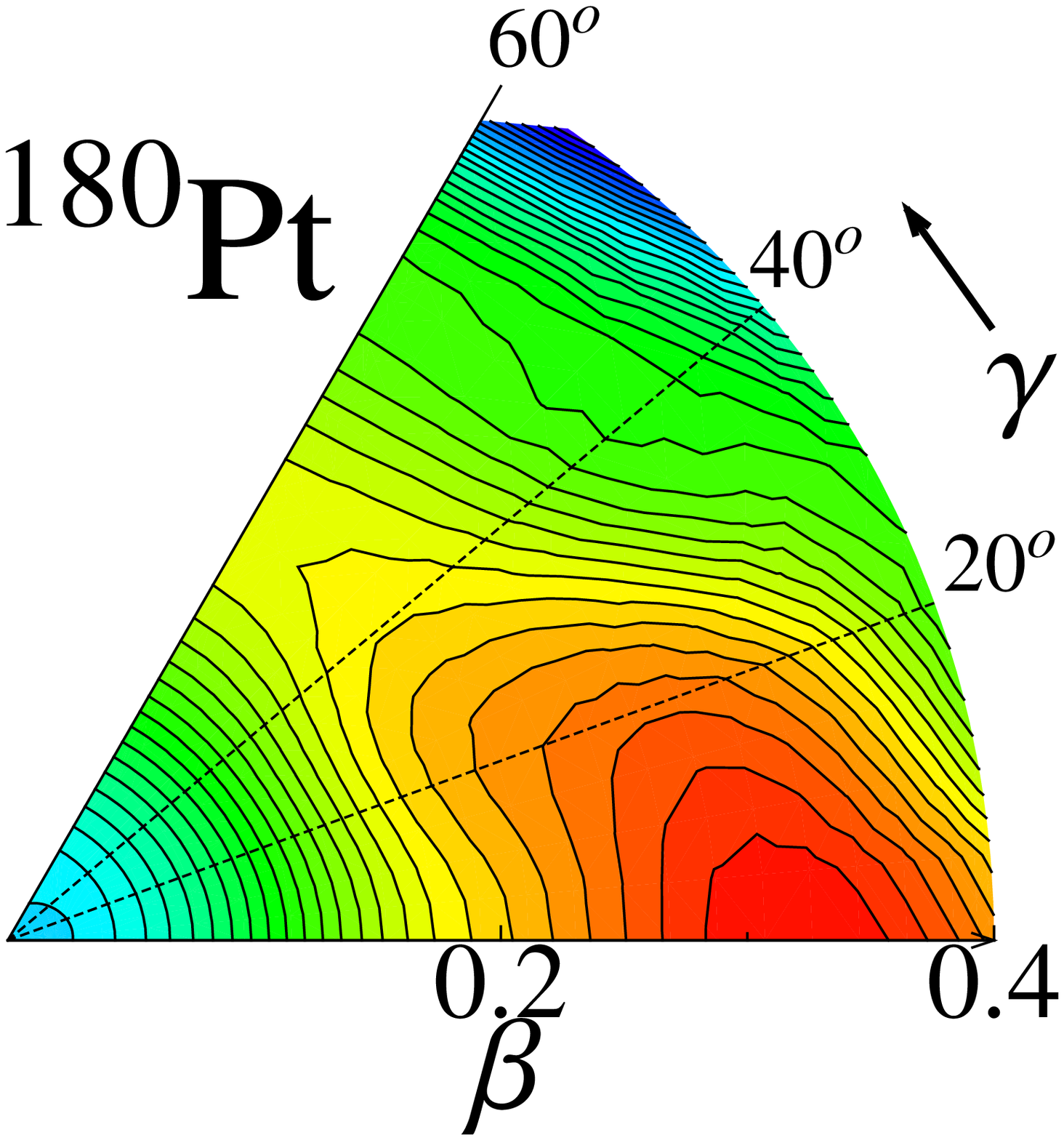}%
\includegraphics[width=0.25\textwidth]{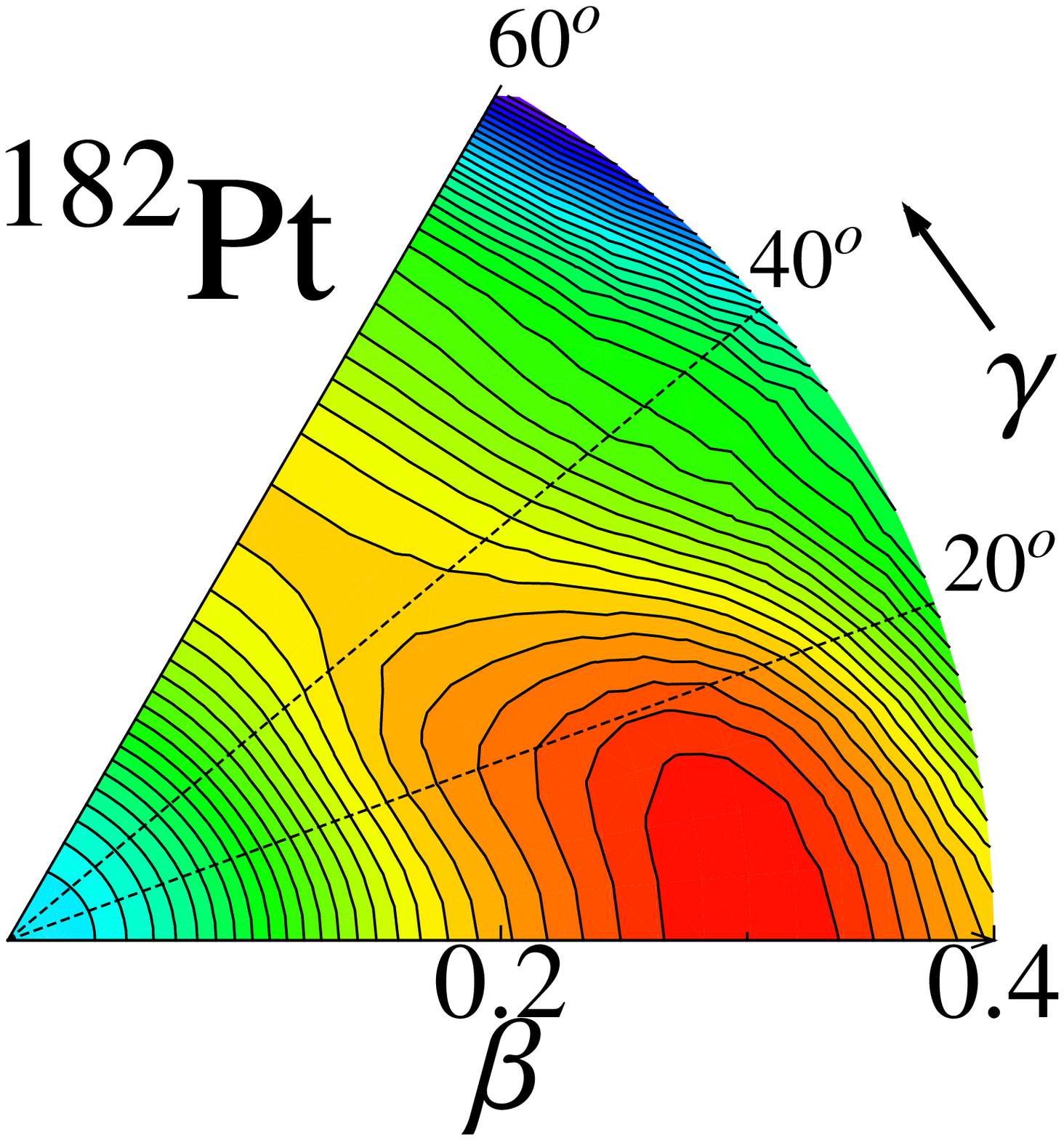}%
\includegraphics[width=0.25\textwidth]{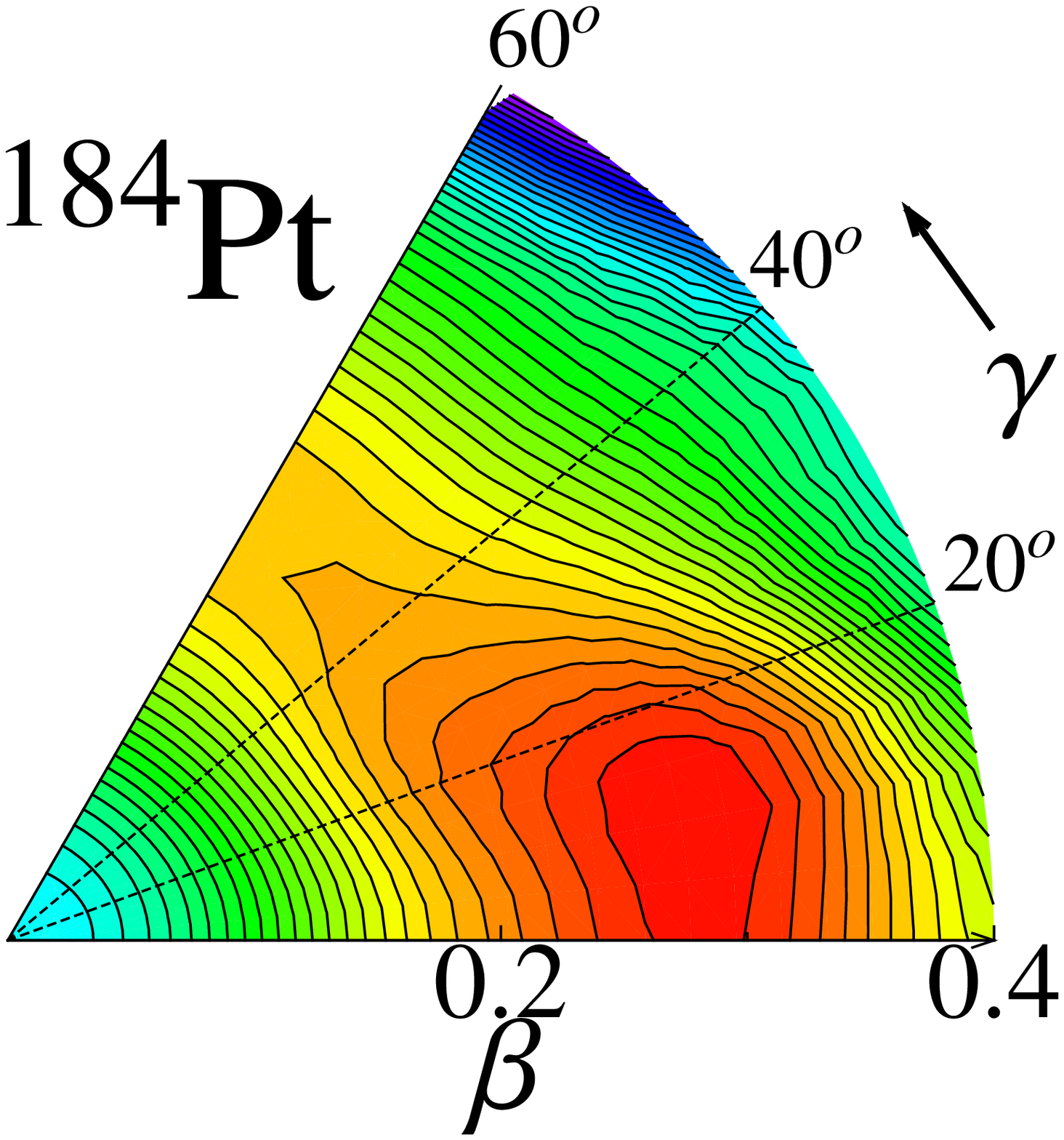}%
\includegraphics[width=0.25\textwidth]{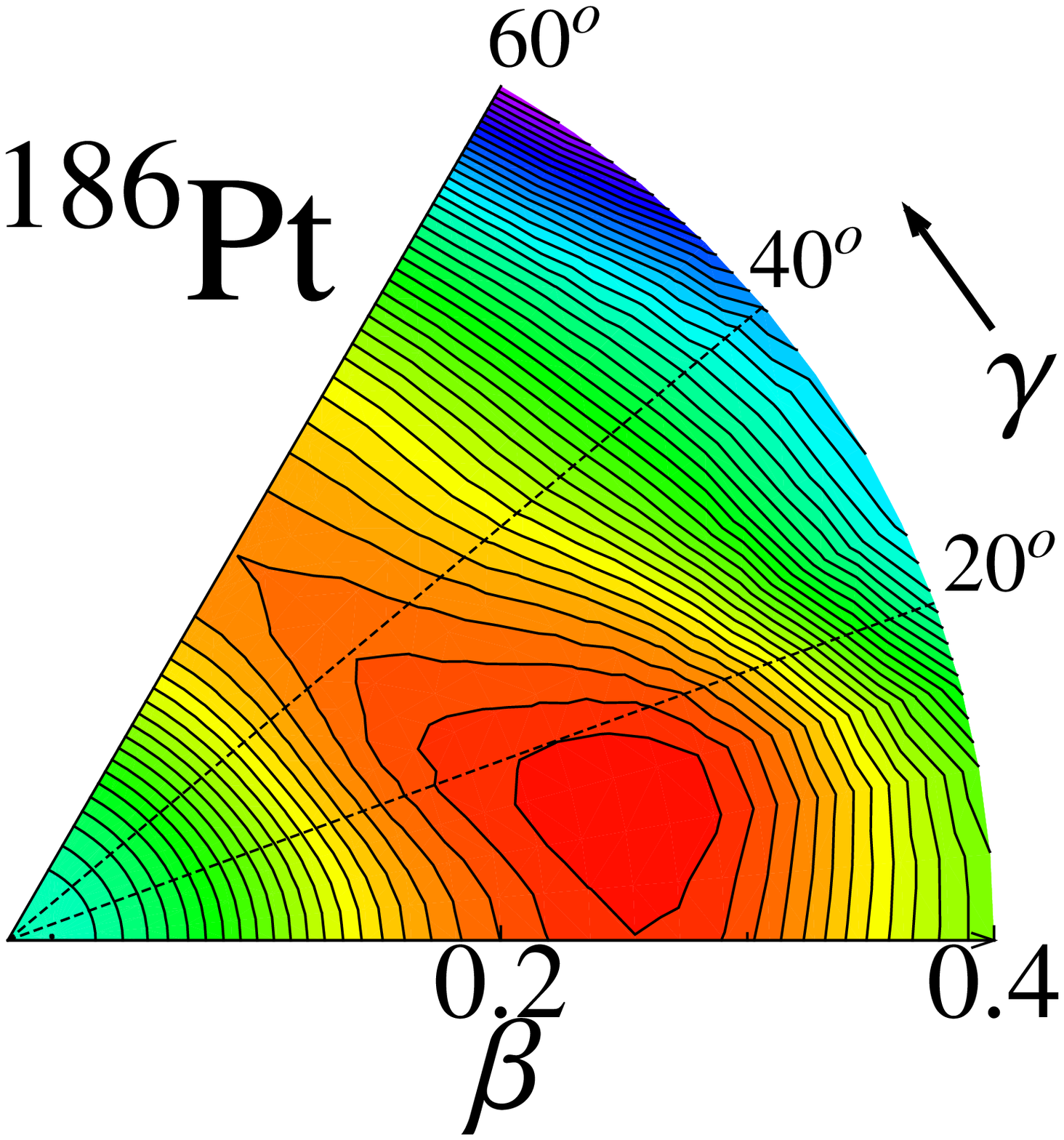}\par
\includegraphics[width=0.25\textwidth]{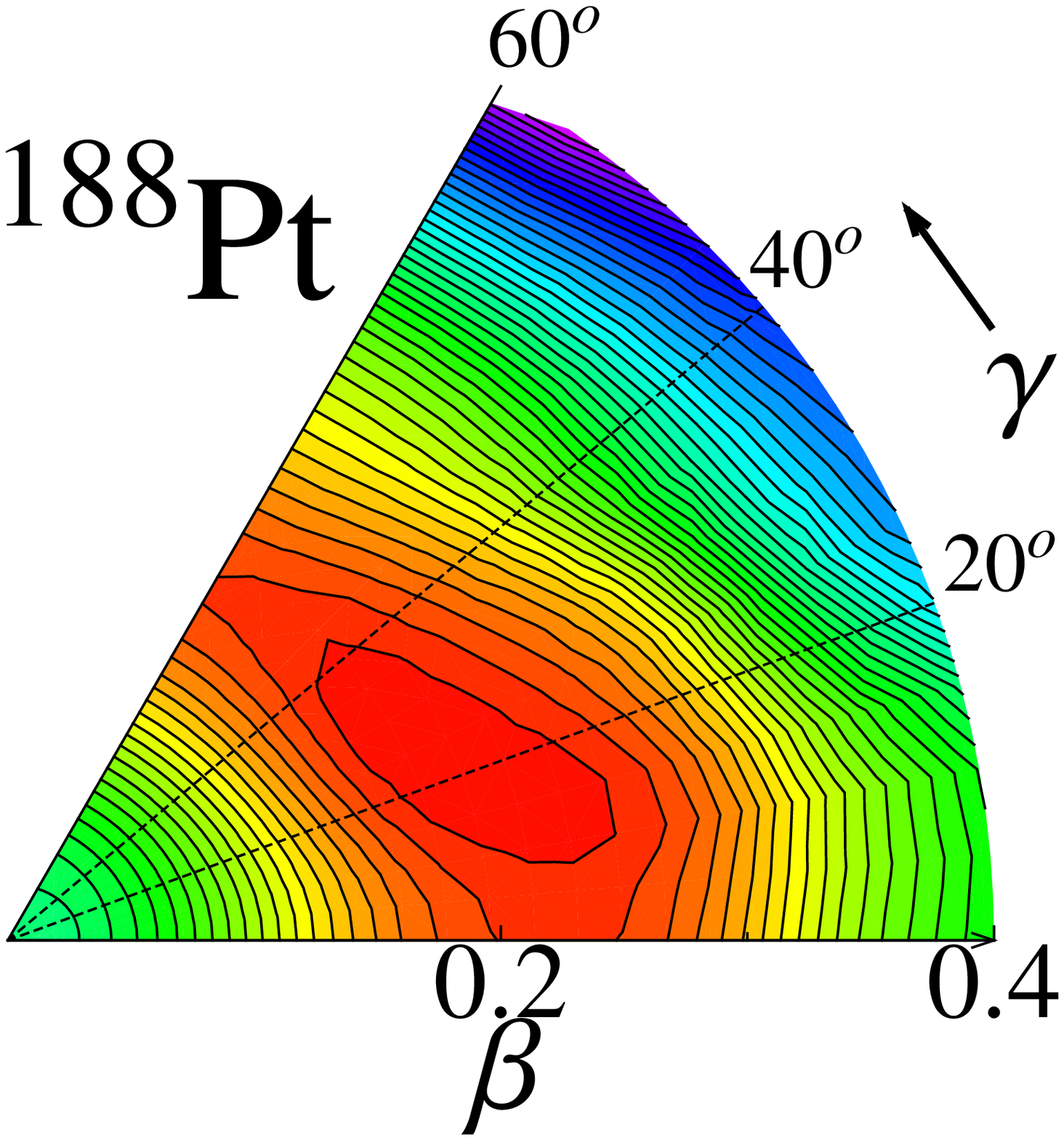}%
\includegraphics[width=0.25\textwidth]{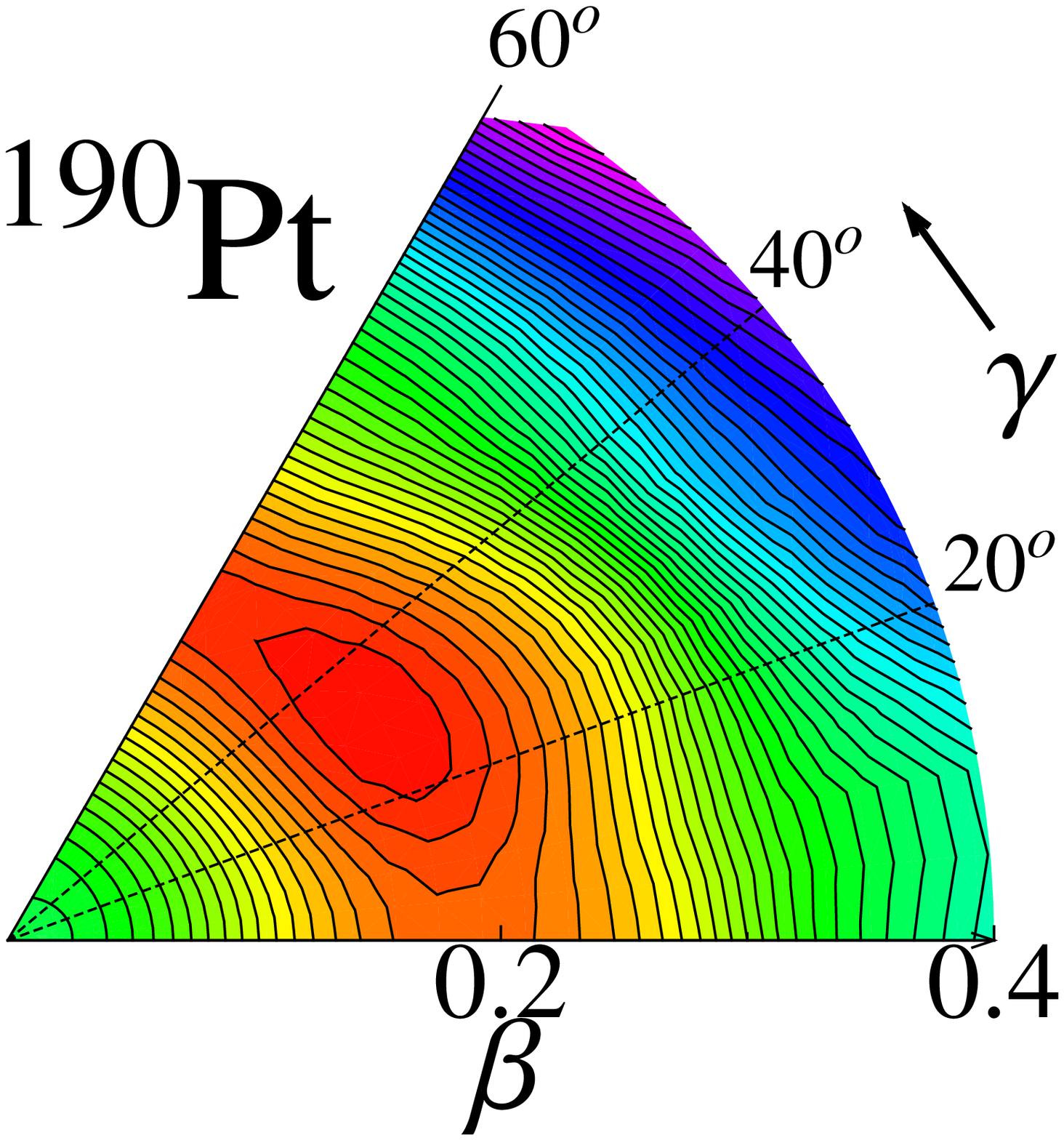}%
\includegraphics[width=0.25\textwidth]{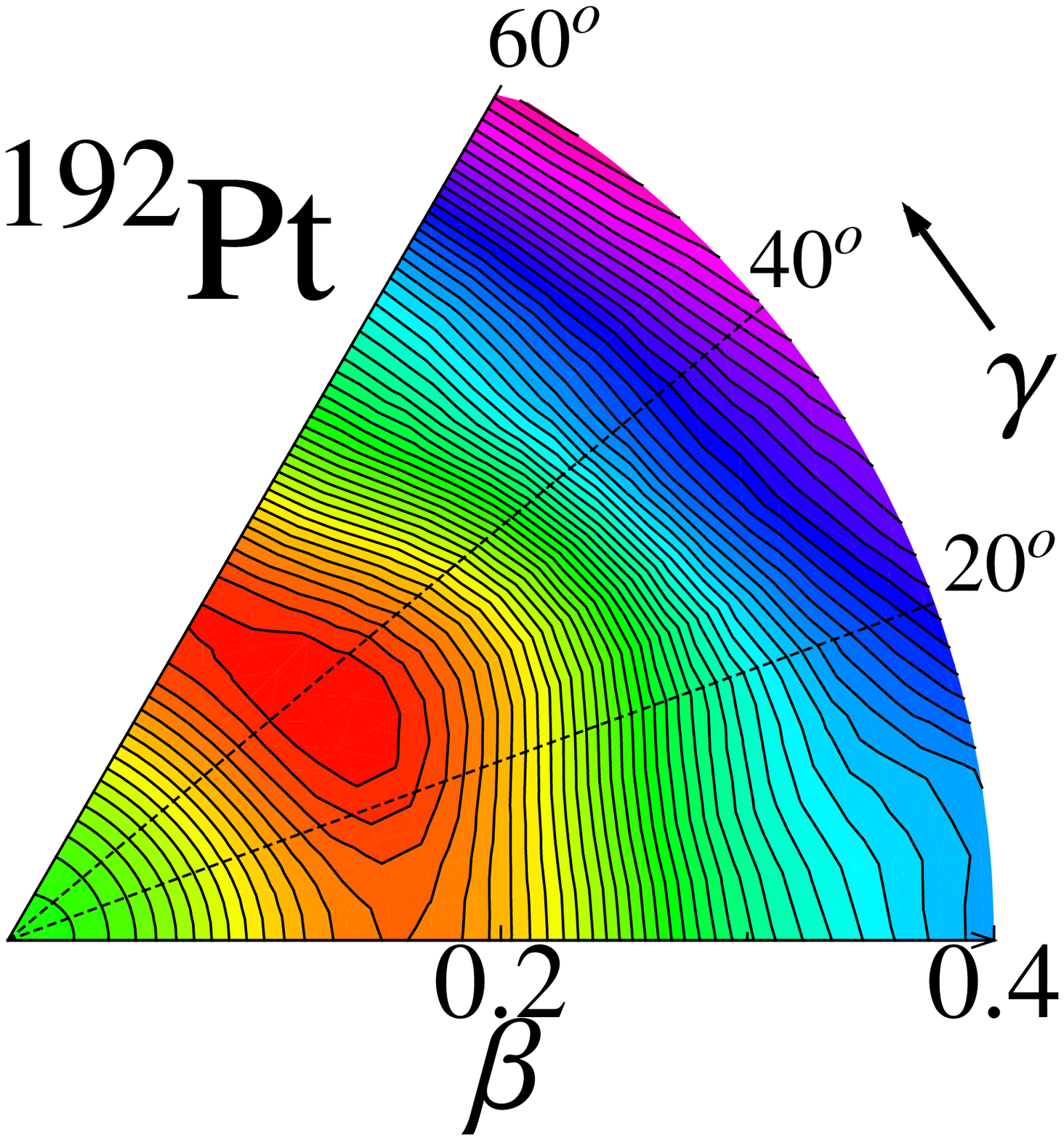}%
\includegraphics[width=0.25\textwidth]{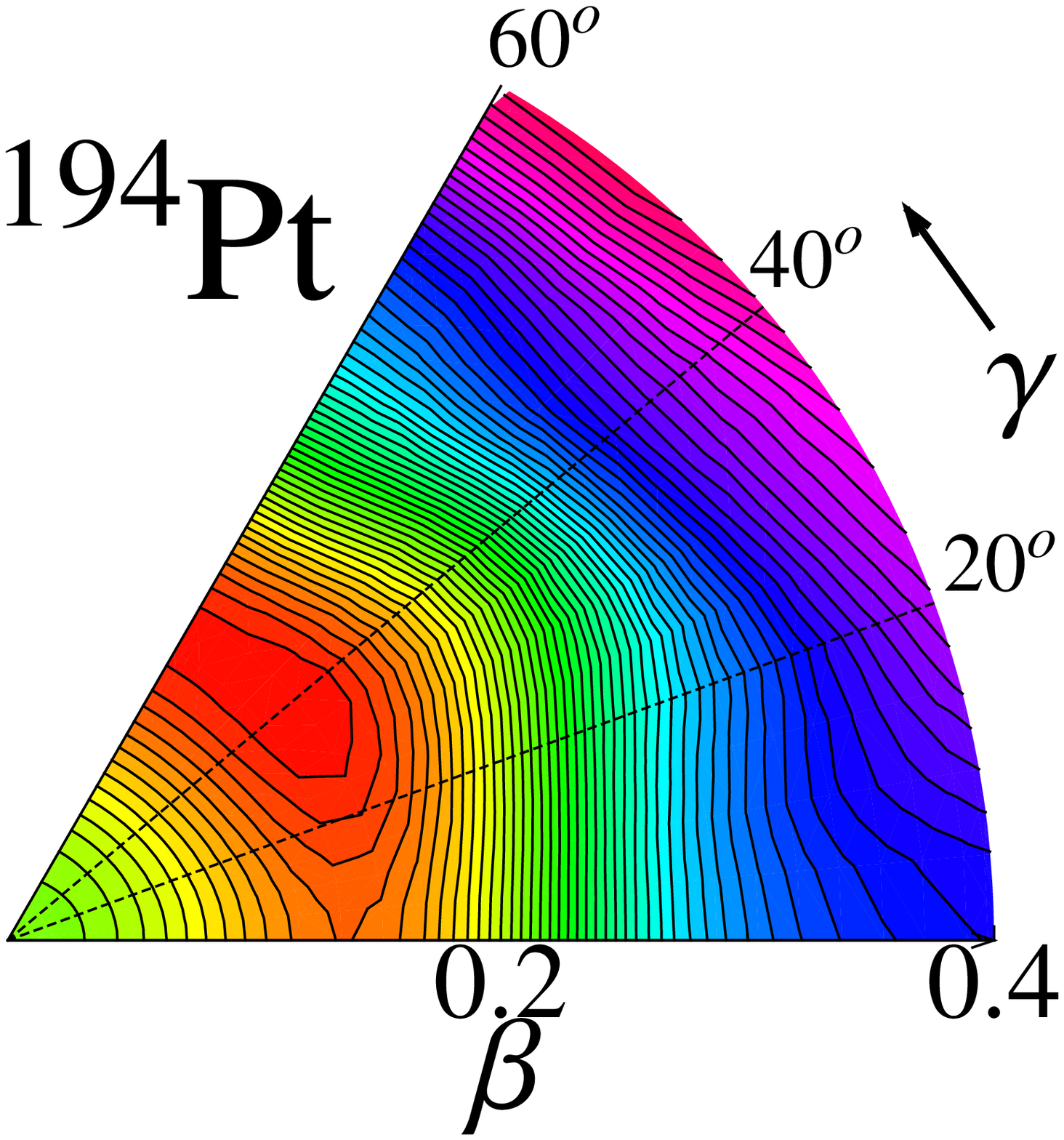}
\caption{(Color online) Hartree-Fock-Bogoliubov contour plots for $^{172-194}$Pt as a function of
  $\beta$ and $\gamma$ (see also Ref.~\cite{rodri10}). 
  The separation between adjacent contour lines amounts to $250$ keV
  and the deepest energy minimum is set to zero, corresponding to 
  the red color, while green corresponds to $\approx 10$ MeV.} 
\label{fig_hfb_ener_surph}
\end{figure*}

In order to carry out a comparison with the 
selfconsistent mean-field HFB calculations using the Gogny-D1S interaction, we show in
Fig.~\ref{fig-HFB-axial-curves} the energy curves along the axial
symmetry axis (see also
Fig.~1 of Ref.~\cite{rodri10}) 
for the same set of Pt isotopes as shown in Fig.~\ref{fig-IBM-axial-curves}. 
%In this figure, the energy is plotted as a function of $\beta$. 
The lightest isotope $^{172}$Pt shows a prolate deformed ground state with
$\beta\approx 0.17$, though less 
deformed than its neighbours, and also presents an oblate local minimum
(see Fig.~\ref{fig_hfb_ener_surph}). 
The isotopes $^{174-184}$Pt present a rather similar structure
with a deep prolate minimum at $\beta\approx 0.3$  and an oblate
local minimum. They display a large
spherical barrier whose height increases from $3.5$ MeV in $^{174}$Pt
to more than $6$ MeV in $^{182}$Pt. 
For these isotopes, we also
observe a reduction in the excitation energy  of the  oblate local minimum
point  with respect to the prolate ground state. On the other hand the
spherical barrier starts to decrease and the prolate and oblate axial
minima become almost degenerate, in particular for $^{188}$Pt.  
From this isotope onwards the ground state turns out to be oblate
while the prolate minimum lies higher in energy.
Finally, the height of the spherical barrier steadily
decreases and the equilibrium value of $\beta$ changes from $\approx
0.2$  in $^{188}$Pt to $\approx 0.13$  in $^{194}$Pt. 

Comparing both approaches, (i) we first of all notice a different
energy 
scale, in particular regarding the height of the spherical barrier.
In the case of the IBM-CM, this height
amounts for $\approx 1.8$ MeV in the largest case, while in the mean-field HFB
approach, this energy becomes as large as $6$ MeV, (ii) on the other hand,
the onset of 
deformation does not happen precisely at the same masses. In particular, $^{172-174}$Pt appear to be almost spherical
in the IBM-CM calculations, while in the
mean-field HFB calculations, a well deformed shape is
obtained, (iii) in spite of the previous differences, the  
similarity between both approaches is  
remarkable for $^{176-188}$Pt, for which the spherical barrier
heights (up to an energy scale), the prolate-oblate energy
differences, and the position of the corresponding minima agree
reasonably well. 
We remark that for $^{188}$Pt, both approaches result in degenerate
prolate and oblate  minima. Finally, for $^{190-194}$Pt  the IBM
produces prolate and oblate minima that are degenerate, 
while mean-field HFB calculations indicate an oblate minimum, even
though the prolate-oblate energy difference is quite small all the
time. It is worth noticing that 
serious differences result between both approaches concerning the energy scale.
A similar situation is observed by Kotila {\it et al.}~\cite{Koti12} where
selfconsistent HF+BCS, mapped IBM, and phenomenological IBM
energy surfaces are compared. The first two present a much larger
energy scale as compared with the phenomenolocical IBM energy surface, which has
been obtained from a fit to the experimental excitation energies
and B(E2) transitions rates. Therefore, these two cases point
towards the existence of quite different energy scales resulting from the microscopic and
phenomenological calculations.

A very convenient way to compare the overall outcome of both
approaches is to plot the value of $\beta$ of the IBM-CM with the one
of HFB as a
function of the mass number.  In Fig.~\ref{fig-comp-beta} (a) we depict the $\beta$ deformation 
parameter provided by the IBM-CM and the one obtained within the Gogny-D1S
HFB framework. In additon, in the same panel (a),
we also plot the $\beta$ value corresponding  to the unperturbed
regular $[N]$ and the intruder $[N+2]$ configurations as
``reference'' values (see also 
Ref.~\cite{Garc12}). As can be seen, there is an overall agreement 
between the mean-field HFB and the IBM approaches. Both approaches point
towards large values of the deformation $\beta$ around neutron mid shell,
though the precise neutron number 
where a sudden onset of deformation arises is different in both cases: the
nucleus  $^{174}$Pt for the HFB calculation and $^{176}$Pt for the IBM-CM calculation,
respectively. 
In Fig.~\ref{fig-comp-beta} (b), we plot the
IBM-CM bandhead energy for both unperturbed structures besides the value of
the ground-state energy (full IBM-CM calculation). Note that these
unperturbed energies correspond to the diagonalization of the IBM
Hamiltonian.
Analyzing these energy curves, one observes a crossing at $^{176-178}$Pt and
at $^{186-188}$Pt. These points delimit the area where the intruder
configuration becomes the ground state. Taking into account that the
intruder configuration corresponds to a more deformed shape than the
regular one (see 
the unperturbed curves in the top panel), it becomes evident why the Pt isotopes
around the mid-shell N=104 (A=182) are more deformed. 
%%%%%%
In the mean-field approximation, deformation results from the spontaneous 
breaking of the rotational symmetry \cite{rodri10,Robl09} in the
considered nuclei. It is quite 
remarkable that only one symmetry-broken mean-field configuration
already accounts for the trend that is otherwise only obtained within
the IBM-CM by invoking the [N+2] boson intruder configurations. On the other
hand, we see how both approximations are complementary with the IBM-CM
providing an intuitive geometrical picture for the onset of deformation
around the neutron mid-shell.

Even though the calculations along the $\beta$ axis (only considering the axially
symmetric shapes) already give a clear view  of the
shape evolution, it is the calculation in the full ($\beta-\gamma$)
plane which provides the more realistic  
``image'' of the nuclear shape. Indeed, we will see that for the heaviest
Pt isotopes the triaxial degree of freedom, $\gamma$, plays a notable
role \cite{rodri10}. 

Before starting with the analysis of the Pt  ($\beta-\gamma$) total
energy contour plots, it is
worth to mention some characteristics of the IBM-CM
parameterization. In order to determine the IBM parameters, as described in \cite{Garc09}, we
imposed a number of constraints 
to reduce the number of free parameters. In particular, we have
considered $\varepsilon_{N+2}$ as well as
$\chi_{N}$ to be equal to zero for the whole isotopic chain. This
latter 
constraint has an immediate consequence on the energy surfaces, {\it
  i.e.}, whenever 
the regular configuration describes the ground state of a given Pt nucleus, the
resulting energy surface will be flat in the $\gamma$ direction, or, 
$\gamma$ unstable. Therefore, in the present IBM-CM approach, the
possibility to  
generate triaxiallity or oblate shapes is absent. In the standard IBM
the triaxiallity is easily accommodated through the inclusion of
three-body terms in the Hamiltonian \cite{Isac81,Heyd84}. In IBM-2
\cite{iach87} and
IBM-CM triaxiallity can be generated with two orthogonal ellipsoids for
protons and neutrons \cite{Levi90} and with the mixing of two
configurations with 
different prolate/oblate character \cite{Hell09}, respectively. Nomura
{\it et al.}~\cite{nomura13,Nomu12} have shown that including a three-boson
interaction within the IBM-2 formulation, treating proton and neutron
bosons explicitely, naturally leads to cover the full ($\beta-\gamma$)
plane.    
This turns out not to form a major drawback of the
present IBM-CM approach when comparing with the mean-field HFB energy surfaces 
for the Pt isotopes. The reason being, as discussed later, 
that the heaviest isotopes, though oblate
in the mean-field HFB approach, are not very far removed from being
$\gamma$ unstable.  

In Fig.~\ref{fig_ibm_ener_surph} we depict the IBM-CM energy surfaces
for the chain of isotopes $^{172-194}$Pt. In $^{172-174}$Pt, we observe
a flat minimum at $\beta=0$, which indeed corresponds to a slightly
deformed 
$\gamma$-unstable minimum as was observed in the corresponding axial
energy plot (see Fig.~\ref{fig-IBM-axial-curves}). In the nucleus $^{176}$Pt,
one observes the onset of a more deformed minimum, resulting in the coexistence 
of a $\gamma$-unstable and a prolate deformed minimum. This isotope is the only
in which two minima coexist. In $^{178-186}$Pt the isotopes present a
well-pronounced prolate minimum around $\beta=0.3$. Finally the nuclei 
$^{188-194}$Pt become less deformed than the medium-mass isotopes and turn
out to be  $\gamma$-unstable. Note that in the present IBM-CM approach, no
genuine triaxial shapes can appear for the Pt isotopes, although, as
mentioned before, triaxiallity can be generated through the use of
more general Hamiltonians.

In Fig.~\ref{fig_hfb_ener_surph} we present the energy contour plots 
computed with the Gogny-D1S interaction for the isotopes
$^{172-194}$Pt (see also Fig.~2 of Ref.~\cite{rodri10}).
In this case, the nuclei $^{172-174}$Pt have a slightly triaxial
shape. On the other hand, the isotope $^{176}$Pt appears as more
deformed than the lighter ones with a prolate shape. Prolate
deformed shapes are also predicted for  $^{178-182}$Pt while a
triaxial shape  develops in $^{184-190}$Pt. In particular, 
for $^{190}$Pt the shape corresponds to $\gamma \approx
30^\circ$. Finally,  $^{192-194}$Pt still present triaxial shapes but
the heavier isotopes come close to exhibit an oblate shape.

In summary, in the IBM-CM approach, the lightest Pt isotopes, {\it
  i.e.}, $^{172-174}$Pt, are slightly deformed. In $^{176}$Pt a prolate
and a $\gamma$-unstable minimum coexist, but quickly, a well deformed
prolate minimum develops in $^{178}$Pt, becoming the most pronounced prolate minimum at the
mid-shell, {\it i.e.}, in $^{182}$Pt with the prolate shape remaining well pronounced up
to $^{186}$Pt. Moving towards heavier mass Pt isotopes, 
$\gamma$-flat energy surfaces start to develop. Indeed, for $^{188}$Pt, a very extended energy surface
develops in the $\gamma$ direction, becoming completely $\gamma$-unstable when reaching
$^{190-194}$Pt. For the set of IBM-CM parameters used in this work no
genuine triaxial shapes can be generated although this can be done using
three-body terms in the Hamiltonian, having orthogonal proton and
neutron ellipsoids in IBM-2 or mixing configurations with
different prolate/oblate character in IBM-CM.
In the mean-field HFB approach, the lightest Pt
isotopes are slightly prolate. Moving to the larger masses, the Pt isotopes become more deformed, 
but at the same time the total energy surface starts to flatten in the $\gamma$ direction.
Passing mid-shell (at N = 104, A = 182), the nuclei become triaxial and the heaviest Pt
isotopes already correspond to oblate shapes. Therefore, the
evolution of the IBM-CM energy surfaces correspond, to a large extent, with the
corresponding total energy surfaces obtained from a mean-field HFB approach. 
The most pronounced difference appears for the heaviest Pt isotopes,
in which the mean-field HFB approach results in triaxial shapes while the IBM-CM
parameterization results in $\gamma$-unstable shapes.      

A similar analysis was carried out by Morales {\it et
al.}~\cite{Mora08} for the even-even Pt nuclei from $A=182$ up to
$A=204$. They started from the set of IBM parameters obtained in the
schematic study of Harder {\it et al.}~\cite{harder97}. The latter
study was performed with the aim of providing a schematic description
of the $^{182-204}$Pt energy spectrum and, indeed, the agreement with
the experimental excitation energies is only qualitative.  In
Ref.~\cite{Mora08} it was shown that the absolute minimum in the
energy surface evolves from a prolate into an oblate shape, finally
turning into a spherical shape with increasing neutron number,
starting at N=104 and ending at the closed shell value of
N=126. Indications for shape coexistence result in the isotopes
$^{182-188}$Pt.  A comparison with the results we derive from the
present IBM-CM Hamiltonian
%and the parametrization fixed through a detailed fitting procedure 
gives an idea about the sensitivity of the energy variations in the
IBM parameters.  

%%%%%%%%%%%%%%%%%%%%%%%%%%%%%%%%%%%%%%%%%%%%%%%%%%%%%%
%%% Nomura 
%%%%%%%%%%%%%%%%%%%%%%%%%%%%%%%%%%%%%%%%%%%%%%%%%%%%%%

It is also interesting to compare the present IBM-CM energy surfaces with
the ones resulting from the mapping of the selfconsistent mean-field
HFB calculations carried out by Nomura {\it et al.}~\cite{nomura11a}. 
In this latter case, the extracted IBM energy surfaces
match very well with the HFB results by obvious reasons and, therefore, 
reproduce the shape evolution from a prolate to
an oblate shape passing through a triaxial region. This is expected
because the mapped Hamiltonian was determined 
through a  mapping of the
HFB total energy surfaces as closely as possible onto the
corresponding IBM mean-field surfaces.
The agreement between
both IBM approaches is reasonable, but once more a large
difference in the energy scale (heigth of the spherical barrier) 
between is observed. The origin of this difference 
deserves further investigation.

%%%%%%%%%%%%%%%%%%%%%%%%%%%%%%%%%%%%%%%%%%%%%%%%%%%%%%%
%Comparison with other HFB calculations%%%%%%%%%%%%%%%%
%%%%%%%%%%%%%%%%%%%%%%%%%%%%%%%%%%%%%%%%%%%%%%%%%%%%%%%

Finally, it is also of interest to compare with previous 
phenomenological mean-field studies.
Calculations by Bengtsson {\it et
al.}~\cite{bengt87}, starting from a deformed Woods-Saxon potential, show that the energy surfaces
obtained for the Pt nuclei turn out to exhibit a more complex behavior when comparing with results obtained 
for the Hg nuclei, covering the region $98 \leq N \leq$ $120$ region.
The prolate minimum is lowest in the mass interval
$178 \leq A \leq 186$, 
%($100 \leq N \leq 108$), 
whereas the oblate minimum becomes lowest for $A=192$ 
%($N=114$) 
and onwards to heavier Pt nuclei. The transition at $A=188, 190$ 
%($N=110, 112$) 
and, at the lower mass side, at $A=176$, 
%($N=98$), 
passes through a $\gamma$-soft energy surface.
Hilberath {\it et al.}~\cite{hilberath92}, who calculated total Routhian surfaces also using a
deformed Woods-Saxon potential, arrive at very much
the same results. Both of these results are largely consistent with the
present IBM-CM matrix coherent-state mean-field results discussed
before. 
Furthermore, calculations within the framework  of the relativistic
mean-field approximation have been carried out by
Fossion {\it et al.}~\cite{fossion06} for the Pt nuclei in the mass
region $184 \leq A \leq 202$. 
%(or, $106 \leq N \leq 124$) are covered. 
The results point out towards a transition from a prolate, as the 
lowest one in $A=186$, towards an oblate minimum, as the lowest one
in $A=188$, with both minima present in the region 
$184 \leq A \leq 192$.
%(or, $106 \leq N \leq 112$). 
Beyond $^{194}$Pt, the energy surface becomes rather flat, evolving towards 
a spherical minimum at $^{200}$Pt and beyond. 
The possibility of triaxial deformation was not considered.

%%%%%%%%%%%%%%%%%%%%%%%%%%%%%%%%%%%%%%%%%%%%%%%%%%%%%%%%%%%%%%%%%%%%%
%%%%%%Conclusions and summary 
%%%%%%%%%%%%%%%%%%%%%%%%%%%%%%%%%%%%%%%%%%%%%%%%%%%%%%%%%%%%%%%%%%%%%
\section{Summary}
\label{sec:summary}

In summary we have carried out a detailed comparison of nuclear total energy surfaces
obtained using two approaches: the IBM-CM, formulated within a laboratory frame, and the
selfconsistent mean-field HFB approach, starting from the Gogny-D1S
interaction, which is an intrinsic frame formulation. The total energy surfaces resulting
from both approaches are qualitatively similar even though they have totally different starting point.
The first one is a more phenomenological approach in which the parameters that determine
the IBM-CM Hamiltonian and, consequently also define the resulting energy surfaces, are
obtained making a careful comparison between the large set of spectroscopic properties and the 
corresponding theoretical results. 
The second approach is a mean-field one based on the Gogny-D1S effective 
interaction whose predicted power all over the nuclear chart has been shown
in previous studies both at the mean-field level and beyond \cite{rodri10}.
Both approaches result
in a consistent description of the nuclear energy landscape which is at the origin of shape 
evolution in the Pt isotopes. The common results point towards the fact that 
the lightest Pt isotopes are slightly deformed and prolate, becoming 
more strongly deformed shapes, while at the same time the potential in the $\gamma$ direction
starts to flatten, developing a triaxial shape once having passed mid shell (at N=104, A=182)
and finally becoming oblate for $^{192-194}$Pt ($\gamma$-unstable in
the IBM-CM case).

Even though total energy surfaces from the above approaches are very similar in structure
for the Pt nuclei discussed in the present paper, 
the final comparison between these approaches has, of course, to 
be made at the level of the nuclear observables: energy spectra,
electromagnetic properties, 
ground-state properties (charge radii), ... 
The IBM, being an algebraic model, has an intrinsic
geometric structure related to the particular Hamiltonian used.  As
discussed in Sect.~\ref{sec:energy}, and amply illustrated in the
present paper, the crucial ingredient is the use of coherent states
resulting  
in energy surfaces that can 
be compared with the results calculated using a selfconsistent
mean-field HFB approach, starting 
from a microscopic basis. It is most interesting to observe a very
similar overall description 
of the energy surfaces describing the long series of Pt isotopes.    
A next step in comparing both approaches will be situated on the level
of beyond mean-field 
calculations (BMF), that allow for the calculation of energy spectra,
electromagnetic properties, ...,
which should  be compared with the results obtained from the IBM-CM
calculations. 
In this way, the effect of the collective inertia, inherent to the GCM,
will be explicitly taken into account.
Comparing a large set of observables, as outlined before, obtained
from a BMF calculation using the Gogny 
interaction, with the corresponding observables calculated using the
mapped IBM Hamiltonian \cite{nomura08}  
will form a major step in order to understand the connection between
both approaches.  
This could be a topic of further study.

%%%%%%%%%%%%%%%%%%%%%%%%%%%%%%%%%%%%%%%%%%%%%%%%%%%%%%%%%%%%%
\section{Acknowledgements}

We thank M. Huyse,
P.~Van Duppen for continuous interest in this research topic and
J.L.~Wood for stimulating discussions in various stages of this work.
Financial support from the ``FWO-Vlaanderen'' (KH and JEGR) and the InterUniversity
Attraction Poles Programme - Belgian State - Federal Office for
Scientific, Technical and Cultural Affairs (IAP Grant No.  P7/12, is
acknowledged.  This work has also been partially supported by the
Spanish Ministerio de Econom\'{\i}a y Competitividad and the European
regional development fund (FEDER) under Project Nos.
FIS2011-28738-C02-02, FPA2012-34694, FIS2012-34479 and by Junta de Andaluc\'{\i}a under Project No.
FQM318, and P07-FQM-02962, as well as by Spanish Consolider-Ingenio 2010
(CPANCSD2007-00042).

\end{document}